%% file: main.tex
\documentclass[aip,jap,reprint]{revtex4-2}

\usepackage{graphicx}
\usepackage{amsmath}
\usepackage{bm}
\usepackage[hidelinks]{hyperref}

\graphicspath{{./}}
\input{numbers.tex}

\newcommand{\Zs}{Z^{*}}
\newcommand{\kB}{k_{\mathrm{B}}}
\newcommand{\MAcm}{MA/cm$^{2}$}
\newcommand{\um}{$\mu$m}

\begin{document}

\title{Solute-Aware Stochastic Korhonen Model of Electromigration in Wide vs Near-Bamboo Al-Cu Lines}

\author{Narahara Chari Dingari}
\email{ndingari@wpi.edu}
\email[Secondary email: ]{ncdingari@gmail.com}
\affiliation{Worcester Polytechnic Institute, Worcester, Massachusetts, USA}
\author{David R. Heskett}
\affiliation{Department of Physics, University of Rhode Island, Kingston, RI 02881, USA}

\date{\today}

\input{sections/abstract}

\maketitle

\input{sections/intro}
\input{sections/model}
\input{sections/methods}
\input{sections/experiment}
\input{sections/results}
\input{sections/comparison}
\input{sections/discussion}
\input{sections/conclusion}

\input{sections/availability}

\bibliography{refs}

\end{document}

%% file: numbers.tex
\newcommand{\PEaGB}{0.60}  %
\newcommand{\PEaB}{1.0}  %
\newcommand{\PDzeroGB}{\ensuremath{3.69\times10^{-10}}}  %
\newcommand{\PDzeroB}{\ensuremath{6.72\times10^{-10}}}  %
\newcommand{\PZAl}{4}  %
\newcommand{\PZCu}{8.6}  %
\newcommand{\PRhoZero}{5.0}  %
\newcommand{\PRhoTzero}{200}  %
\newcommand{\PAlphaRho}{\ensuremath{1.888\times10^{-3}}}  %
\newcommand{\PB}{50}  %
\newcommand{\POmega}{\ensuremath{1.0\times10^{-29}}}  %
\newcommand{\PSigCrit}{150}  %
\newcommand{\PSigCC}{350}  %
\newcommand{\PEaCu}{0.76}  %
\newcommand{\PDzeroCuSI}{\ensuremath{5.01\times10^{-7}}}  %
\newcommand{\PBeta}{\ensuremath{1.147\times10^{4}}}  %
\newcommand{\PDfifty}{1.7}  %
\newcommand{\PSigLn}{0.28}  %
\newcommand{\PSigKappa}{0.28}  %
\newcommand{\PWideW}{10}  %
\newcommand{\PNarrowW}{3}  %
\newcommand{\PWDNB}{1.0}  %
\newcommand{\PReliefRate}{1000}  %
\newcommand{\PReliefWidth}{1}  %
\newcommand{\PEaCuHost}{0.80}  %
\newcommand{\PDzeroCuHost}{\ensuremath{1.3\times10^{-9}}}  %
\newcommand{\PZCuHost}{1}  %
\newcommand{\PRhoCu}{\ensuremath{2.25\times10^{-8}}}  %
\newcommand{\PBCu}{28}  %
\newcommand{\POmegaCu}{\ensuremath{1.18\times10^{-29}}}  %
\newcommand{\PSigCritCu}{41}  %
\newcommand{\PDzeroMnSI}{\ensuremath{8.0\times10^{-6}}}  %
\newcommand{\PEaMn}{1.778}  %
\newcommand{\PDeltaEaMn}{0.2}  %
\newcommand{\PBetaMn}{\ensuremath{9.26\times10^{4}}}  %
\newcommand{\PBetaMnLow}{\ensuremath{1.1\times10^{4}}}  %
\newcommand{\PBetaMnHigh}{\ensuremath{9.3\times10^{4}}}  %
\newcommand{\PDGBT}{\ensuremath{1.5\times10^{-16}}}  %
\newcommand{\PDBT}{\ensuremath{1.5\times10^{-20}}}  %
\newcommand{\PDCuT}{\ensuremath{4.0\times10^{-15}}}  %
\newcommand{\PEllThree}{1.6}  %
\newcommand{\PGLThree}{26.9}  %
\newcommand{\PTnucEstThree}{0.046}  %
\newcommand{\PPBambooNarrow}{0.021}  %
\newcommand{\PWDNarrow}{1.76}  %
\newcommand{\PWDWide}{5.9}  %
\newcommand{\PCzeroHalf}{\ensuremath{2.13\times10^{-3}}}  %
\newcommand{\PCzeroOne}{\ensuremath{4.271\times10^{-3}}}  %
\newcommand{\PBlockHalf}{25.4}  %
\newcommand{\PBlockTwo}{99.5}  %
\newcommand{\PDCuOverDGB}{27}  %
\newcommand{\PCsolWt}{0.10}  %
\newcommand{\PCsolAt}{\ensuremath{4.2\times10^{-4}}}  %
\newcommand{\PCsolQR}{4031}  %
\newcommand{\PCsolWtOneFifty}{0.037}  %
\newcommand{\PCsolWtTwoFifty}{0.23}  %
\newcommand{\PCsolWtThreeHundred}{0.44}  %
\newcommand{\PCapWidth}{5.0}  %
\newcommand{\PMobFracHalf}{0.20}  %
\newcommand{\PMobFracTwo}{0.050}  %
\newcommand{\NVoneMaxErr}{\ensuremath{1.6\times10^{-7}}}  %
\newcommand{\NVoneMaxErrTau}{0.002}  %
\newcommand{\NVoneCells}{1600}  %
\newcommand{\NConvOrderMin}{1.97}  %
\newcommand{\NConvOrderMax}{2.00}  %
\newcommand{\NConvNmin}{100}
\newcommand{\NConvNmax}{1600}
\newcommand{\NConvErrFine}{\ensuremath{6.5\times10^{-9}}}  %
\newcommand{\NConvErrCoarse}{\ensuremath{1.6\times10^{-6}}}  %
\newcommand{\NSteadyRuns}{100}  %
\newcommand{\NSteadyFailed}{0}  %
\newcommand{\NSteadyDsig}{0.0037}  %
\newcommand{\NRegSample}{200}  %
\newcommand{\NGCMaps}{100}  %
\newcommand{\NGCRuns}{700}  %
\newcommand{\NGCMedMin}{0.06}  %
\newcommand{\NGCMedMax}{4.0}  %
\newcommand{\NGCTfiftyMax}{3.7}  %
\newcommand{\NGCFlip}{0}  %
\newcommand{\NGCSiteMin}{8}  %
\newcommand{\NGCSiteMax}{24}  %
\newcommand{\NEoneNMin}{2.00}  %
\newcommand{\NEoneNMax}{2.05}  %
\newcommand{\NEoneEaCuZeroMin}{0.634}  %
\newcommand{\NEoneEaCuZeroMax}{0.635}  %
\newcommand{\NEoneEaCuHalfMin}{0.688}  %
\newcommand{\NEoneEaCuHalfMax}{0.728}  %
\newcommand{\NEoneEaCuHalfSresZero}{0.728}
\newcommand{\NEoneEaCuHalfSresMax}{0.688}
\newcommand{\NEoneRuns}{144}  %
\newcommand{\PSresMax}{75}  %
\newcommand{\NEoneTCuZeroSZero}{0.046}  %
\newcommand{\NEoneTCuHalfSZero}{1.0}  %
\newcommand{\NEoneCuFactor}{22}  %
\newcommand{\NImmWideResCuZeroJOne}{0.701}  %
\newcommand{\NImmWideResCuZeroJTwo}{0.131}  %
\newcommand{\NTWideResCuZeroJThree}{1.68}  %
\newcommand{\NSigWideResCuZeroJThree}{1.1}  %
\newcommand{\NImmWideResCuZeroJThree}{0.007}  %
\newcommand{\NTWideBlkCuZeroJThree}{0.0471}  %
\newcommand{\NSigWideBlkCuZeroJThree}{0.20}  %
\newcommand{\NTWideResCuHalfJThree}{3.38}  %
\newcommand{\NSigWideResCuHalfJThree}{0.18}  %
\newcommand{\NImmWideResCuHalfJThree}{0.000}  %
\newcommand{\NTWideBlkCuHalfJThree}{1.03}  %
\newcommand{\NSigWideBlkCuHalfJThree}{0.20}  %
\newcommand{\NTWideResCuTwoJThree}{11.2}  %
\newcommand{\NSigWideResCuTwoJThree}{0.20}  %
\newcommand{\NImmWideResCuTwoJThree}{0.000}  %
\newcommand{\NTWideBlkCuTwoJTwo}{9.16}  %
\newcommand{\NTWideBlkCuTwoJThree}{4.06}  %
\newcommand{\NSigWideBlkCuTwoJThree}{0.20}  %
\newcommand{\NTNarrowResCuZeroJThree}{0.0519}  %
\newcommand{\NSigNarrowResCuZeroJThree}{0.70}  %
\newcommand{\NImmNarrowResCuZeroJThree}{0.000}  %
\newcommand{\NTNarrowBlkCuZeroJThree}{0.0424}  %
\newcommand{\NSigNarrowBlkCuZeroJThree}{0.15}  %
\newcommand{\NTNarrowResCuHalfJThree}{1.05}  %
\newcommand{\NSigNarrowResCuHalfJThree}{0.29}  %
\newcommand{\NImmNarrowResCuHalfJThree}{0.000}  %
\newcommand{\NTNarrowBlkCuHalfJThree}{0.926}  %
\newcommand{\NSigNarrowBlkCuHalfJThree}{0.15}  %
\newcommand{\NTNarrowBlkCuHalfJFive}{0.328}  %
\newcommand{\NTNarrowResCuTwoJThree}{4.24}  %
\newcommand{\NSigNarrowResCuTwoJThree}{0.26}  %
\newcommand{\NImmNarrowResCuTwoJThree}{0.000}  %
\newcommand{\NTNarrowBlkCuTwoJThree}{3.71}  %
\newcommand{\NSigNarrowBlkCuTwoJThree}{0.16}  %
\newcommand{\NTNBResCuZeroJThree}{0.0433}  %
\newcommand{\NSigNBResCuZeroJThree}{0.11}  %
\newcommand{\NImmNBResCuZeroJThree}{0.000}  %
\newcommand{\NTNBBlkCuZeroJThree}{0.0432}  %
\newcommand{\NSigNBBlkCuZeroJThree}{0.11}  %
\newcommand{\NTNBResCuHalfJThree}{0.928}  %
\newcommand{\NSigNBResCuHalfJThree}{0.12}  %
\newcommand{\NImmNBResCuHalfJThree}{0.000}  %
\newcommand{\NTNBBlkCuHalfJThree}{0.924}  %
\newcommand{\NSigNBBlkCuHalfJThree}{0.12}  %
\newcommand{\NTNBResCuTwoJThree}{3.76}  %
\newcommand{\NSigNBResCuTwoJThree}{0.13}  %
\newcommand{\NImmNBResCuTwoJThree}{0.000}  %
\newcommand{\NTNBBlkCuTwoJThree}{3.74}  %
\newcommand{\NSigNBBlkCuTwoJThree}{0.13}  %
\newcommand{\NNBJOneKMMin}{1489}  %
\newcommand{\NNBJOneKMMax}{3752}  %
\newcommand{\NNBJOneLognMin}{100}  %
\newcommand{\NNBJOneLognMax}{815}  %
\newcommand{\NNBJOneSigMin}{2.0}  %
\newcommand{\NNBJOneSigMax}{4.9}  %
\newcommand{\NSigBlkWNMin}{0.096}  %
\newcommand{\NSigBlkWNMax}{0.26}  %
\newcommand{\NSigBlkNBJgeMin}{0.11}  %
\newcommand{\NSigBlkNBJgeMax}{0.13}  %
\newcommand{\NCuFacWideBlkHalf}{21.9}  %
\newcommand{\NCuFacWideBlkTwo}{86.2}  %
\newcommand{\NCuFacWideResHalf}{2.01}  %
\newcommand{\NCuFacWideResTwo}{6.66}  %
\newcommand{\NCuFacNarrowBlkHalf}{21.9}  %
\newcommand{\NCuFacNarrowBlkTwo}{87.5}  %
\newcommand{\NCuFacNBBlkHalf}{21.4}  %
\newcommand{\NCuFacNBBlkTwo}{86.5}  %
\newcommand{\NWNRatioBlkMin}{1.10}  %
\newcommand{\NWNRatioBlkMax}{1.12}  %
\newcommand{\NThesisRatioBlk}{28}  %
\newcommand{\NThesisRatioRes}{73}  %
\newcommand{\NGapWideBlk}{16}  %
\newcommand{\NGapNarrowBlk}{805}  %
\newcommand{\NImmMaxNarrow}{0.024}
\newcommand{\NImmMaxNB}{0.011}

\newcommand{\NEthreeRuns}{108000}  %
\newcommand{\NEthreeMaps}{1000}  %

\newcommand{\NEthreeSteady}{874}

\newcommand{\NCathNarrowBlk}{48}  %
\newcommand{\NLtTenNarrowBlk}{54}  %
\newcommand{\NLtHalfNarrowBlk}{75}  %
\newcommand{\NGtNineNarrowBlk}{4.3}  %
\newcommand{\NCathNBBlk}{11}  %
\newcommand{\NLtHalfNBBlk}{55}  %
\newcommand{\NCathWideBlk}{100}  %
\newcommand{\NTdivShare}{83.4}  %
\newcommand{\NResHalfMin}{49}  %
\newcommand{\NResHalfMax}{51}  %
\newcommand{\NHillockWideRes}{5}  %

\newcommand{\NTzeroOneRatioNarrowBlk}{0.66}  %
\newcommand{\NTzeroOneRatioNarrowBlkLo}{0.63}  %
\newcommand{\NTzeroOneRatioNarrowBlkHi}{0.68}  %
\newcommand{\NTzeroOneRatioWideBlk}{0.60}  %
\newcommand{\NTzeroOneRatioWideBlkLo}{0.58}  %
\newcommand{\NTzeroOneRatioWideBlkHi}{0.64}  %
\newcommand{\NTzeroOneRatioWideRes}{0.60}  %
\newcommand{\NTzeroOneRatioWideResLo}{0.56}  %
\newcommand{\NTzeroOneRatioWideResHi}{0.64}  %
\newcommand{\NNoReliefRuns}{4800}
\newcommand{\NNoReliefMaps}{200}
\newcommand{\NSigMinNoRelief}{-0.474}  %
\newcommand{\NNoReliefBlkMaxRel}{0}  %
\newcommand{\NNoReliefImmWideResJTwo}{0}
\newcommand{\NReliefImmWideResJTwoSame}{0}  %
\newcommand{\NBnWideBlkMin}{2.006}  %
\newcommand{\NBnWideBlkMax}{2.019}  %
\newcommand{\NBnNarrowBlkMin}{2.025}  %
\newcommand{\NBnNarrowBlkMax}{2.047}  %
\newcommand{\NBnNBJgeMin}{2.12}  %
\newcommand{\NBnNBJgeMax}{2.25}  %
\newcommand{\NBnNBJOneMin}{4.90}  %
\newcommand{\NBnNBJOneMax}{8.09}  %
\newcommand{\NBnWideResCuZeroJFourSix}{3.56}  %
\newcommand{\NBnWideResCuHalfJOneThree}{2.37}  %
\newcommand{\NBnWideResCuHalfJTwoFive}{2.23}  %
\newcommand{\NBnNarrowResCuZeroJTwoFive}{2.05}  %
\newcommand{\NBnNarrowResCuHalfJTwoFive}{1.99}  %
\newcommand{\NBnNarrowResCuTwoJTwoFive}{1.99}  %
\newcommand{\NBnNBBlkCuZeroJOneThree}{7.35}  %
\newcommand{\NBnNBBlkCuZeroJTwoFive}{2.25}  %
\newcommand{\NBnNBBlkCuZeroJFourSix}{2.15}  %
\newcommand{\NBnNBBlkCuHalfJTwoFive}{2.23}  %
\newcommand{\NBnNBBlkCuTwoJOneThree}{4.90}  %
\newcommand{\NBnNBBlkCuTwoJTwoFive}{2.23}  %
\newcommand{\NBnGated}{1}  %
\newcommand{\NBnWideResCuZeroJTwo}{5.27}
\newcommand{\NBnWideResCuZeroJTwoImm}{0.13}
\newcommand{\NBnCIHalfMax}{0.052}  %
\newcommand{\NBnCuShiftMax}{0.01}  %
\newcommand{\NImmGate}{0.2}  %
\newcommand{\NEtwoMaps}{20}

\newcommand{\NAnodeNBBlkCuHalf}{1.00}  %
\newcommand{\NAnodeWideBlkCuHalf}{1.03}  %
\newcommand{\NAnodeWideBlkCuOne}{1.03}  %
\newcommand{\NAnodeWideBlkCuTwo}{1.03}  %
\newcommand{\NAnodeWideResCuHalf}{1.11}  %
\newcommand{\NAnodeWideResCuOne}{1.10}  %
\newcommand{\NAnodeWideResCuTwo}{1.09}  %
\newcommand{\NTincWideRes}{0.77}  %
\newcommand{\NTincFinBlk}{98}  %
\newcommand{\NEtwoBlkRuns}{180}
\newcommand{\NSeedPeakWide}{8.3}  %
\newcommand{\NSeedPeakXWide}{0.67}  %
\newcommand{\NSeedTnucWide}{3605}  %
\newcommand{\NCuMaxNucMin}{8.2}  %
\newcommand{\NCuMaxNucMax}{16}  %
\newcommand{\NCuMaxNucPct}{11}  %
\newcommand{\NSiteCuMinWideResMin}{0.043}  %
\newcommand{\NSiteCuMinWideResMax}{0.16}  %
\newcommand{\NSiteCuMaxWideResMin}{2.2}  %
\newcommand{\NSiteCuMaxWideResMax}{2.6}  %
\newcommand{\NJLcrit}{936}  %
\newcommand{\NLcrit}{4.68}  %
\newcommand{\NEfourLImm}{4}  %
\newcommand{\NEfourLFail}{5}  %
\newcommand{\NEfourMaps}{200}

\newcommand{\NEfourCapLImm}{4}  %
\newcommand{\NEfourCapLFail}{5}  %
\newcommand{\NEfourCapMaps}{20}  %
\newcommand{\NEfourCapRuns}{160}  %
\newcommand{\NCuMnTZero}{22.5}  %
\newcommand{\NCuMnTHalf}{\ensuremath{1.2\times10^{4}}}  %
\newcommand{\NCuMnTTwo}{\ensuremath{4.8\times10^{4}}}  %
\newcommand{\NCuMnFacHalf}{536}
\newcommand{\NCuMnFacTwo}{2136}
\newcommand{\NCuMnImmWide}{0.73}  %
\newcommand{\NCuMnFailWide}{27}  %
\newcommand{\NCuMnImmNarrow}{0.70}  %
\newcommand{\NCuMnImmNB}{0.99}  %

\newcommand{\NEsixRuns}{10800}
\newcommand{\NEsixMaps}{100}

\newcommand{\NEsevenRuns}{14{,}400}  %
\newcommand{\NEsevenMaps}{200}  %
\newcommand{\NEsevenTendStops}{0}  %
\newcommand{\NSevenRatioBlkWNCuHalfMin}{1.16}  %
\newcommand{\NSevenRatioBlkWNCuHalfMax}{1.20}  %
\newcommand{\NSevenRatioNBCuHalfMin}{1.18}  %
\newcommand{\NSevenRatioNBCuHalfMax}{1.24}  %

\newcommand{\NSevenRatioBlkWNCuTwoMin}{1.15}  %
\newcommand{\NSevenRatioBlkWNCuTwoMax}{1.19}  %
\newcommand{\NSevenRatioNBCuTwoMin}{1.18}  %
\newcommand{\NSevenRatioNBCuTwoMax}{1.24}  %

\newcommand{\NSevenTImmWideBlkCuTwoJThree}{4.70}  %
\newcommand{\NSevenRatioWideResMin}{4.19}  %
\newcommand{\NSevenRatioWideResMax}{8.40}  %
\newcommand{\NSevenRatioNBJOneMin}{1.0}  %
\newcommand{\NSevenRatioNBJOneMax}{2.6}  %
\newcommand{\NSevenStaticCuHalf}{25.4}  %
\newcommand{\NSevenStaticCuTwo}{99.5}  %
\newcommand{\NSevenImmImmCuHalfJOne}{0.72}  %
\newcommand{\NSevenImmImmCuHalfJTwo}{0.15}  %
\newcommand{\NSevenNImmWideResCuHalfJFourSix}{3.55}  %
\newcommand{\NSevenNImmNBBlkCuHalfJTwoFive}{2.26}  %
\newcommand{\NSevenNImmNBBlkCuTwoJTwoFive}{2.28}  %
\newcommand{\NTenRuns}{28{,}800}  %
\newcommand{\NTenRatioTotBlkWNCuHalfMin}{1.33}  %
\newcommand{\NTenRatioTotBlkWNCuHalfMax}{1.38}  %
\newcommand{\NTenRatioTotNBCuHalfMin}{1.1}  %
\newcommand{\NTenRatioTotNBCuHalfMax}{1.3}  %
\newcommand{\NTenImmTotCuHalfJOne}{0}  %
\newcommand{\NTenImmMobCuHalfJOne}{0.99}  %
\newcommand{\NTenImmTotCuHalfJTwo}{0}  %
\newcommand{\NTenImmMobCuHalfJTwo}{0.46}  %
\newcommand{\NTenRatioTotBlkWNCuTwoMin}{2.24}  %
\newcommand{\NTenRatioTotBlkWNCuTwoMax}{2.36}  %
\newcommand{\NTenRatioTotNBCuTwoMin}{1.2}  %
\newcommand{\NTenRatioTotNBCuTwoMax}{2.0}  %
\newcommand{\NTenTTotWideBlkCuTwoJThree}{4.09}  %
\newcommand{\NTenTMobWideBlkCuTwoJThree}{1.73}  %
\newcommand{\NTenRatioDisBlkWNCuHalfMin}{0.363}  %
\newcommand{\NTenRatioDisBlkWNCuHalfMax}{0.371}  %
\newcommand{\NTenImmDisCuHalfJOne}{0.90}  %
\newcommand{\NTenTDisWideBlkCuHalfJThree}{0.277}  %
\newcommand{\NTenRatioDisBlkWNCuTwoMin}{0.153}  %
\newcommand{\NTenRatioDisBlkWNCuTwoMax}{0.161}  %
\newcommand{\NTenImmDisCuTwoJOne}{0.86}  %
\newcommand{\NTenTDisWideBlkCuTwoJThree}{0.278}  %
\newcommand{\NTenNTotWideResCuHalfJFourSix}{2.21}  %
\newcommand{\NTenNMobWideResCuHalfJFourSix}{3.01}  %
\newcommand{\NTenNTotWideResCuTwoJFourSix}{2.26}  %
\newcommand{\NTenNMobWideResCuTwoJFourSix}{3.43}  %
\newcommand{\NTenNTotNBBlkCuHalfJTwoFive}{2.21}  %
\newcommand{\NTenNMobNBBlkCuHalfJTwoFive}{2.37}  %
\newcommand{\NTenNTotNBBlkCuTwoJTwoFive}{2.23}  %
\newcommand{\NTenNMobNBBlkCuTwoJTwoFive}{2.75}  %
\newcommand{\NTenDnBlkWNMax}{0.02}  %
\newcommand{\NElevenRuns}{43{,}200}  %
\newcommand{\NElevenImmPartMax}{0}  %
\newcommand{\NElevenImmZeroJOneMin}{0.86}  %
\newcommand{\NElevenImmZeroJOneMax}{0.90}  %
\newcommand{\NElevenRatioQuarterCuHalfMin}{0.48}  %
\newcommand{\NElevenRatioQuarterCuHalfMax}{0.49}  %
\newcommand{\NElevenRatioQuarterCuTwoMin}{0.33}  %
\newcommand{\NElevenRatioQuarterCuTwoMax}{0.34}  %
\newcommand{\NElevenRatioHalfCuHalfMin}{0.67}  %
\newcommand{\NElevenRatioHalfCuHalfMax}{0.68}  %
\newcommand{\NElevenRatioHalfCuTwoMin}{0.58}  %
\newcommand{\NElevenRatioHalfCuTwoMax}{0.59}  %
\newcommand{\NElevenRatioThreeQCuHalfMin}{0.84}  %
\newcommand{\NElevenRatioThreeQCuHalfMax}{0.85}  %
\newcommand{\NElevenRatioThreeQCuTwoMin}{0.80}  %
\newcommand{\NElevenRatioThreeQCuTwoMax}{0.81}  %
\newcommand{\NElevenNWideResZeroMin}{3.4}  %
\newcommand{\NElevenNWideResZeroMax}{3.5}  %
\newcommand{\NElevenNWideResQuarterMin}{2.6}  %
\newcommand{\NElevenNWideResQuarterMax}{2.7}  %
\newcommand{\NElevenNNBMin}{2.2}  %
\newcommand{\NElevenCuRatioZeroMin}{1.00}  %
\newcommand{\NElevenCuRatioZeroMax}{1.12}  %
\newcommand{\NElevenCuRatioPartMin}{1.9}  %
\newcommand{\NElevenCuRatioPartMax}{4.0}  %
\newcommand{\NTwelveMaps}{200}  %
\newcommand{\NTwelveWideAnode}{1.73}  %
\newcommand{\NTwelveWideAnodeLo}{1.61}  %
\newcommand{\NTwelveWideAnodeHi}{1.89}  %
\newcommand{\NTwelveWideContrast}{7.0}  %
\newcommand{\NTwelveWideContrastLo}{4.5}  %
\newcommand{\NTwelveWideContrastHi}{17}  %
\newcommand{\NTwelveWideMidPct}{66}  %
\newcommand{\NTwelveWideAnodeMaxPct}{100}  %
\newcommand{\NTwelveNarrowDepUm}{56}  %
\newcommand{\NTwelveNarrowDepUmLo}{8.7}  %
\newcommand{\NTwelveNarrowDepUmHi}{211}  %
\newcommand{\NTwelveNarrowPeakCathPct}{4}  %
\newcommand{\NTwelveNarrowPeakHalfPct}{22}  %
\newcommand{\NTwelveWideAnodeSixH}{1.03}  %
\newcommand{\NTwelveNarrowPeakCathPctTTF}{3}  %
\newcommand{\NTwelveNarrowPeakCathPctTFiftyC}{2}  %
\newcommand{\NStepRuns}{1200}  %
\newcommand{\NStepFallPct}{100}  %
\newcommand{\NStepFallRandPct}{55}  %
\newcommand{\NStepSignedMed}{0.12}  %
\newcommand{\NStepRankPct}{76}  %
\newcommand{\NThirteenBaseMin}{1.15}  %
\newcommand{\NThirteenBaseMax}{1.19}  %
\newcommand{\NThirteenDFastMin}{6.25}  %
\newcommand{\NThirteenDFastMax}{10.5}  %
\newcommand{\NThirteenSolTwoMin}{1.39}  %
\newcommand{\NThirteenSolTwoMax}{1.49}  %
\newcommand{\NThirteenZMin}{1.12}  %
\newcommand{\NThirteenZHiMax}{1.22}  %
\newcommand{\NThirteenResDSlowMax}{1.5}  %
\newcommand{\NThirteenResBaseMin}{12.3}  %
\newcommand{\NThirteenResBaseMax}{14.4}  %
\newcommand{\NThirteenMaps}{100}  %
\newcommand{\NBootWidthMin}{0.22}  %
\newcommand{\NBootWidthMax}{1.4}  %
\newcommand{\NBootB}{400}  %
\newcommand{\NGridTwoMaxT}{10}  %
\newcommand{\NGridTwoClass}{0}  %
\newcommand{\NEightSigCritHalf}{0.275}  %
\newcommand{\NEightSigCritTwo}{3.30}  %
\newcommand{\NEightSigCritFour}{9.59}  %
\newcommand{\NEightSigCritGPa}{19}  %
\newcommand{\NEightDzeroTenth}{3.83}  %
\newcommand{\NEightDzeroTen}{0.116}  %
\newcommand{\NEightEaMinus}{0.334}  %
\newcommand{\NEightEaPlus}{2.26}  %
\newcommand{\NEightBetaHalf}{0.572}  %
\newcommand{\NEightBetaTwo}{1.62}  %
\newcommand{\NEightZLow}{1.66}  %
\newcommand{\NEightZHigh}{0.666}  %
\newcommand{\NEightCombined}{26}  %
\newcommand{\NEightCombinedGPa}{156}  %
\newcommand{\NEightJDiffMax}{1}  %
\newcommand{\NEightSingleMax}{3.8}  %
\newcommand{\NEightSingleMin}{0.12}  %
\newcommand{\NEightBaseJTwo}{2.27}  %
\newcommand{\NEightBaseJFive}{0.363}  %
\newcommand{\NEightRuns}{34}
\newcommand{\NEnineRuns}{64{,}800}  %
\newcommand{\NEnineMaps}{200}  %
\newcommand{\NNineEaBlkMin}{0.63}  %
\newcommand{\NNineEaBlkMax}{0.79}  %
\newcommand{\NNineNBnLowMin}{4.8}  %
\newcommand{\NNineNBnLowMax}{7.6}  %
\newcommand{\NNineAccTlo}{250}  %
\newcommand{\NNineAccThi}{300}  %
\newcommand{\NNinenAccBlkWNMin}{2.03}  %
\newcommand{\NNinenAccBlkWNMax}{2.06}  %
\newcommand{\NNineRatioNarrowResMin}{0.77}  %
\newcommand{\NNineRatioNarrowResMax}{1.8}  %
\newcommand{\NNineNBJoneFacMin}{22}  %
\newcommand{\NNineNBJoneFacMax}{7{,}300}  %
\newcommand{\NNineRatioNBJtwoMin}{0.66}  %
\newcommand{\NNineRatioNBJtwoMax}{1.4}  %
\newcommand{\NNinenAccNBMin}{1.94}  %
\newcommand{\NNinenAccNBMax}{2.16}  %
\newcommand{\NNinenAccWideResMin}{1.94}  %
\newcommand{\NNinenAccWideResMax}{3.30}  %
\newcommand{\NNineWideResImmUse}{2}  %
\newcommand{\NNineWideResUse}{12}  %
\newcommand{\NNineRatioWideResMin}{0.072}  %
\newcommand{\NNineRatioWideResMax}{0.47}  %
\newcommand{\NNineRatioMax}{2.33}  %
\newcommand{\NNineRatioBlkWNCuTwoOneFiftyMin}{2.1}  %
\newcommand{\NNineRatioBlkWNCuTwoOneFiftyMax}{2.3}  %
\newcommand{\NNineRatioBlkWNCuTwoTwoHundredMin}{1.2}  %
\newcommand{\NNineRatioBlkWNCuTwoTwoHundredMax}{1.3}  %
\newcommand{\NNineRatioBlkWNLowCuMax}{1.1}  %
\newcommand{\NNineOverTol}{10}  %
\newcommand{\NNineOverCount}{12}  %
\newcommand{\NNineCount}{72}  %
\newcommand{\NNineEaLocCuTwoLo}{0.68}  %
\newcommand{\NNineEaLocCuTwoHi}{0.89}  %
\newcommand{\NNineEaLocCuZeroLo}{0.63}  %
\newcommand{\NNineEaLocCuZeroHi}{0.64}  %
\newcommand{\NEthreeMobRuns}{72{,}000}  %
\newcommand{\NTotalRuns}{351{,}628}  %

%% file: sections/abstract.tex
\begin{abstract}
We couple an evolving Cu field to the one-dimensional Korhonen stress equation on random grain maps for wide (10~$\mu$m), narrow (3~$\mu$m) and near-bamboo Al-Cu lines. Earlier random-microstructure simulations fixed the Cu effect, and most continuum solute models used prescribed structures with all Cu mobile; we join the two and let only dissolved Cu move, the excess staying as immobile precipitate. With 10$^{3}$ maps per condition at 200~$^{\circ}$C, we report nucleation times and sites, immortal fractions and Black's-law parameters. In cathode-limited lines Cu depletion shortens the median nucleation time \NSevenRatioBlkWNCuTwoMin- to \NSevenRatioBlkWNCuHalfMax-fold relative to immobile Cu for the cited Cu diffusivity; treating all Cu as mobile overstates this a further \NTenRatioTotBlkWNCuTwoMin- to \NTenRatioTotBlkWNCuTwoMax-fold at 2~wt\%; the exponent stays near 2. In wide lines with reservoir ends, Cu steps left by the immobile precipitate become the failure sites, removing their Cu-free immortality if a quarter or more of the precipitate blocks Al diffusion. With Cu above the solvus the lifetime is not Arrhenius, as the dissolved fraction rises with temperature: a Black's-law fit at 250 to 300~$^{\circ}$C overestimates the median nucleation time at 150~$^{\circ}$C up to \NNineRatioMax-fold at 2~wt\% Cu with an exponent near 2. At the test durations the model qualitatively matches the cathode-side damage in narrow lines and the Cu build-up and mid-line peaks in wide lines but not the cathode-side Cu peaks or longer lives of narrow lines.
\end{abstract}

%% file: sections/intro.tex
\section{Introduction}
\label{sec:intro}

Electromigration (EM), the drift of metal atoms under the momentum transfer from conduction electrons, limits the lifetime of interconnects in integrated circuits.\cite{black1969,black1969b,ho2022} In a line whose ends block atomic flux, the drift builds a mechanical stress gradient that opposes it. Korhonen and co-workers wrote this balance as a diffusion equation for the hydrostatic stress,\cite{korhonen1993} and Clement and Thompson developed it into a model of stress evolution in confined lines.\cite{clement1995} A void nucleates when the tensile stress reaches a critical value. Below the Blech product $(jL)_{\mathrm{crit}}$ of current density and line length the steady-state stress stays below that value and the line is immortal.\cite{blech1976} For nucleation-limited failure the Korhonen equation gives a lifetime proportional to $j^{-2}$, the exponent $n = 2$ of Black's law.\cite{shatzkes1986,lloyd2007}

Small additions of Cu raise the EM lifetime of Al lines.\cite{ames1970,lloyd1990,hu1993} Copper segregates to grain boundaries and slows Al grain-boundary diffusion.\cite{rosenberg1972,liu1999} Copper itself drifts toward the anode,\cite{blech1977,kao2001} so the region near the cathode is depleted of Cu first. Aluminum transport, and damage, begins in the depleted region after an incubation time.\cite{hu1993,kim1993,proost2000} In bamboo lines the benefit of Cu is smaller and the spread of failure times is narrower.\cite{theiss1997} Several models couple Cu and stress. Korhonen \emph{et al.} combined stress, electric potential and concentration gradients as driving forces.\cite{korhonen1996} Lloyd and Clement derived the $j^{-2}$ scaling for a Cu-depletion front moving away from a blocking boundary.\cite{lloyd1996} Park, Andleigh and Thompson coupled Al and Cu chemical potentials to stress evolution in bamboo and near-bamboo lines with prescribed structures, and found $n = 2$ for nucleation and $n = 1$ for growth in both pure Al and Al-Cu.\cite{park1999} Kao \emph{et al.} reproduced measured Cu and stress profiles with a finite-element model in which Al is blocked where Cu exceeds a threshold,\cite{kao2003} and recent work on Al(Cu) lines continues this approach.\cite{gordineer2024,wang2024} Kim, Morris and Lee treated the precipitates explicitly: they modeled the growth of a precipitate-free zone at the cathode end and related the incubation time to the time for that zone to exceed the Blech length.\cite{kim1997}

A second line of work treats the microstructure as random. Polygranular clusters in near-bamboo lines set the failure statistics.\cite{korhonen1993stat,joo1994,lloyd1991,dwyer2004} Knowlton, Clement and Thompson generated large populations of realistic lines with a grain-growth simulator and solved a Korhonen-type stress model on them; they obtained failure-time distributions, the transition between polygranular and transgranular failure, and the current-density and temperature scaling of near-bamboo lines.\cite{knowlton1997,knowlton1998,knowltonthesis} Korhonen \emph{et al.} solved the stress equation on Monte Carlo grain structures and extracted the apparent activation energy and current-density exponent.\cite{korhonen1999,korhonen2000} Dwyer computed nucleation-time distributions from random microstructures and the Korhonen equation.\cite{dwyer2008,dwyer2010} Mechanism maps show how the failure mode changes with $j$ and $L$ in pure and alloyed lines,\cite{andleigh1999} and stress models for Cu interconnects now range from microstructure-resolved\cite{sukharev2007,saleh2023} to full-chip\cite{huang2016,sun2020} models.

To our knowledge, these two lines of work have not been joined. In Knowlton's simulations Cu is a fixed reduction of the cluster diffusivity, set so that the ratio of cluster to bamboo diffusivity is 200 instead of $10^{4}$ for pure Al.\cite{knowltonthesis} The solute models above use one prescribed structure at a time, and most of the continuum ones let all the Cu move. As a result, the joint effect of an evolving Cu field and a random grain structure on the lifetime distribution, on the nucleation site, and on the apparent Black's-law parameters has not been computed.

The question is not confined to the Al-Cu lines of the 1990s. Electromigration of Al interconnects at 200 to 300~$^{\circ}$C is one of the limits on high-temperature electronics, which space exploration, geothermal, avionic and automotive systems need.\cite{pradhan2024} In SiC power devices run at junction temperatures of 300 to 450~$^{\circ}$C, Al electrodes in source contact holes fail by electromigration with an apparent exponent $n = 3.0$ to 3.4, and lifetimes are extrapolated over two decades of current density with that exponent held fixed.\cite{tanimoto2010} In Cu damascene lines a solute again acts on grain-boundary transport. In Cu(Mn) lines, Cu-solute binding reduces Cu grain-boundary migration and Cu depletion at the cathode is preceded by an incubation period in which Mn is depleted first, as in Al(2~at.\%~Cu), and in CoWP-capped Cu lines the activation energy falls from 2.2 to 0.72~eV as the grain structure goes from near-bamboo to polycrystalline,\cite{hu2014} the same three ingredients as in Al-Cu. At advanced nodes, interconnects, including the buried power rails of backside power delivery, are bamboo or near-bamboo with only a few grains across the width, and their grain structure changes the electromigration driving forces.\cite{jiang2026} Finally, physics-based analysis based on the Korhonen equation is being developed to replace the rule-based checks built on Black's law and the Blech criterion, but has not yet reached practice; in it, process variation enters as a spread of the activation energy, and lifetimes measured in accelerated tests are carried to use conditions with the same current-density exponent.\cite{sapatnekar2026} Whether a mobile solute and a random microstructure change $n$ and $E_a$ between test and use conditions bears directly on that step.

This paper addresses that gap. The step is incremental, and its contributions are four.

\emph{(1) A coupled solute-stress model on random microstructures.} We couple an evolving Cu concentration field $c(x,t)$, with drift, diffusion and depletion, to the Korhonen stress equation on populations of random grain-segment maps. Here the local Al diffusivity depends on the local Cu content, and only Cu below its solubility moves; the excess is held as immobile precipitate. Knowlton\cite{knowltonthesis,knowlton1997} solved Korhonen stress on simulated near-bamboo microstructures, but represented Cu only as a fixed reduction of the cluster diffusivity ($D_c/D_b = 200$), and left Cu coupling as future work. Park, Andleigh and Thompson\cite{park1999} coupled Cu and Al transport to stress on prescribed line structures. Korhonen \emph{et al.}\cite{korhonen2000} used Monte Carlo grain structures without solute. Our model joins these two lines of work in one 1D solver. We test whether the evolving field matters by repeating the population runs with the Cu held immobile and with all Cu mobile, the assumption of earlier solute models. It does, in a way that depends on where the lines fail: Cu depletion before nucleation shortens the median nucleation time of cathode-dominated lines without changing $n$, and in wide lines with reservoir ends the Cu steps left by the immobile precipitate become the failure sites. Treating all Cu as mobile overstates the first effect and reverses the second.

\emph{(2) Lifetime distributions along a Cu-content axis, for near-bamboo and polygranular lines.} Using populations of $10^{3}$ maps per condition, we report nucleation-time distributions and immortal fraction as functions of initial Cu content and the Blech product $jL$, for both near-bamboo and polygranular ($W \gg D_{50}$) lines. Knowlton's simulations established the near-bamboo trends with $W/D_{50}$, $L/D_{50}$, $j$ and $T$, including bimodal failure statistics, using about 10 lines per population and two fixed diffusivity ratios (pure Al and Al-Cu).\cite{knowltonthesis} We add a Cu-content axis (0, 0.5 and 2~wt\%), the polygranular class, and population sizes large enough to resolve early-failure ($t_{01}$) tails.

\emph{(3) A test of whether solute depletion shifts the apparent Black parameters.} We ask whether a Cu-depletion incubation stage changes the apparent current-density exponent $n$ and activation energy $E_a$ that would be extracted from simulated populations. Previous work found $n = 2$ for nucleation-limited failure in Al-Cu.\cite{park1999,lloyd1996} Knowlton showed that $n$ and $E_a$ depart from their diffusion values across mechanism transitions and near immortality, without solute.\cite{knowltonthesis} Korhonen \emph{et al.} extracted apparent $n$ and $E_a$ from microstructure populations.\cite{korhonen2000} We separate the solute contribution from the mechanism-transition contribution by comparing runs with capped, immobile, fully mobile and no Cu, and report the answer whether or not $n$ departs from 2. We then add a temperature axis (150 to 300~$^{\circ}$C), across which the Cu solubility changes about tenfold, and test how well a Black's-law fit at accelerated conditions predicts the simulated lifetimes at lower $T$ and $j$.

\emph{(4) Damage-location and Cu-profile signatures compared with measurements.} We predict where voids nucleate along the line, at the cathode, at interior cluster ends or toward the anode, together with the accompanying Cu profiles. We compare these predictions qualitatively with optical, atomic force and scanning electron microscopy and energy-dispersive spectroscopy of Al-Cu lines stressed at 200~$^{\circ}$C.\cite{dingari2007} Knowlton reported only whether the stress maximum lay at cluster ends or line ends. He used signed tensile and compressive criteria to compare lifetimes and activation energies, not to locate damage.\cite{knowltonthesis} Neither Knowlton nor Park \emph{et al.} reported failure-position statistics or Cu profiles.

Several ingredients are not new. Cu-stress coupling,\cite{korhonen1996,lloyd1996,park1999,kao2003} Korhonen stress on random grain maps,\cite{korhonen1993stat,knowlton1997,korhonen2000,dwyer2008} (the last for Cu lines) simulated lifetime distributions versus line width, bimodal statistics in near-bamboo lines, and the failure of Black's-law scaling across mechanism transitions\cite{knowlton1997,knowlton1998,knowltonthesis} are all published. Signed tensile and compressive thresholds follow Knowlton.\cite{knowltonthesis} The model is one-dimensional and predicts void nucleation only, not void growth or resistance change. An open, validated solver is part of the work but is not claimed as a contribution in itself. We also apply the same solver to Cu(Mn) lines, to show that it is not specific to Al-Cu; that case is not a contribution either.

Section~\ref{sec:model} describes the model and its parameters, Sec.~\ref{sec:methods} the numerical method, its validation and the statistics, and Sec.~\ref{sec:experiment} the experiment. Section~\ref{sec:results} reports the simulations, and Sec.~\ref{sec:comparison} the measurements and their comparison with the simulations. Section~\ref{sec:discussion} discusses the limits of the model.

%% file: sections/model.tex
\section{Model}
\label{sec:model}

\subsection{Stress and solute transport}

The line occupies $0 \le x \le L$, with the cathode at $x = 0$. Electrons, and with them the Al atoms, move toward $+x$. The state of the line is the hydrostatic stress $\sigma(x,t)$, tensile positive, and the total Cu atomic fraction $c(x,t)$, dissolved and precipitated. The stress follows the Korhonen equation\cite{korhonen1993}
\begin{equation}
\frac{\partial \sigma}{\partial t} = \frac{\partial}{\partial x}\left[\kappa_{\mathrm{eff}}(x,c)\left(\frac{\partial \sigma}{\partial x} + G\right)\right],
\qquad G = \frac{e \Zs_{\mathrm{Al}} \rho j}{\Omega},
\label{eq:stress}
\end{equation}
with $\kappa = D_a B \Omega/(\kB T)$, where $D_a$ is the effective atomic diffusivity, $B$ the effective modulus, $\Omega$ the atomic volume and $\rho(T) = \rho_0[1 + \alpha_\rho (T - T_0)]$ the resistivity. The quantity $\kappa(\partial_x\sigma + G)$ is proportional to the atomic flux. With blocking ends and constant $\kappa$ the steady state is $\sigma = \sigma_{\mathrm{res}} + G(L/2 - x)$, tensile at the cathode. We do not include the stress dependence of the diffusivity used by Knowlton.\cite{knowltonthesis}

Only dissolved Cu moves. At 200~$^{\circ}$C the solubility of Cu in Al is $c_{\mathrm{sol}} = \PCsolWt$~wt\% (\PCsolAt),\cite{knowltonthesis,murray1985} well below the Cu content of the lines, so most of the Cu is in $\theta$-Al$_{2}$Cu precipitates. We take the dissolved fraction as $c_m = \min(c, c_{\mathrm{sol}})$, with the excess held as immobile precipitate in local equilibrium with it; precipitates dissolve where dissolved Cu is carried away and grow where it arrives, without a rate limit. The Cu fraction then obeys
\begin{equation}
\frac{\partial c}{\partial t} = \frac{\partial}{\partial x}\left[D_s(x)\left(\frac{\partial c_m}{\partial x} - v c_m\right)\right],
\qquad v = \frac{|\Zs_{\mathrm{Cu}}| e \rho j}{\kB T},
\label{eq:solute}
\end{equation}
where $D_s$ is the Cu diffusivity. For the solver the minimum is replaced by a $C^1$ blend over $\pm\PCapWidth\%$ of $c_{\mathrm{sol}}$. At other temperatures $c_{\mathrm{sol}}$ follows the solvus, $c_{\mathrm{sol}}(T) \propto \exp(-\PCsolQR\,\mathrm{K}/T)$, fitted to Knowlton's solvus points\cite{knowltonthesis} (\PCsolWtOneFifty, \PCsolWtTwoFifty{} and \PCsolWtThreeHundred~wt\% at 150, 250 and 300~$^{\circ}$C). Two bounds are also run: all Cu mobile ($c_m = c$) and no Cu mobile (Sec.~\ref{sec:res:cutransport}). The measured effective charge of Cu in Al is negative,\cite{kao2001} so Cu drifts with the electron wind toward the anode. Copper blocks Al grain-boundary diffusion. On grain-boundary material we use
\begin{equation}
\kappa_{\mathrm{eff}} = \frac{\kappa}{1 + \beta c},
\label{eq:block}
\end{equation}
with $c$ the total Cu, so Cu in grain-boundary precipitates blocks as well as dissolved Cu. This is a linear trapping form of the kind used by Rosenberg\cite{rosenberg1972} and by Park \emph{et al.};\cite{park1999} on bamboo material $\kappa_{\mathrm{eff}} = \kappa$, because Knowlton's bamboo diffusivity does not depend on Cu.\cite{knowltonthesis} The constant $\beta$ is calibrated, without reference to our simulations, from Knowlton's cluster diffusivities at 200~$^{\circ}$C, $1.5\times10^{-16}$~m$^{2}$/s for pure Al and $3\times10^{-18}$~m$^{2}$/s for Al-1~wt\% Cu.\cite{knowltonthesis} The ratio of 50 at $c = \PCzeroOne$ gives $\beta = 49/c = \PBeta$. Knowlton chose the Al-Cu value to fit failure times, so $\beta$ inherits that calibration and assumes that blocking is linear in $c$ at every Cu content.

On grain-boundary material $D_a = D_{0,\mathrm{gb}} \exp(-E_{a,\mathrm{gb}}/\kB T)$ and $D_s = D_{0,\mathrm{Cu}} \exp(-E_{a,\mathrm{Cu}}/\kB T)$. On bamboo material $D_a = D_{0,b} \exp(-E_{a,b}/\kB T)$, the interface-dominated value, and $D_s$ is the grain-boundary Cu diffusivity scaled by the same ratio $D_b/D_{\mathrm{gb}}$ as Al. At 200~$^{\circ}$C these give $D_{\mathrm{gb}} = \PDGBT$~m$^{2}$/s, $D_b = \PDBT$~m$^{2}$/s and $D_s = \PDCuT$~m$^{2}$/s on grain boundaries, so Cu diffuses \PDCuOverDGB{} times faster than Al. Table~\ref{tab:params} lists every input with its source. The initial state is uniform, $\sigma = \sigma_{\mathrm{res}}$ and $c = c_0$, with $c_0$ the atomic fraction of the nominal Cu content ($c_0 = \PCzeroHalf$ at 0.5~wt\%).

\subsection{Boundary conditions, relief and events}

Each end of the line is either \emph{blocking}, with zero atomic flux, $\kappa(\partial_x\sigma + G) = 0$, as at a W stud or via, or a \emph{reservoir}, with $\sigma = \sigma_{\mathrm{res}}$, as at a large pad. Knowlton used both.\cite{knowltonthesis} The Cu flux is zero at both ends in every run, so the Cu inventory of the line is conserved and pads or vias do not exchange Cu with the line. This is an assumption, discussed in Sec.~\ref{sec:discussion}.

A void nucleates when the tensile stress first reaches $\sigma_{\mathrm{crit}} = \PSigCrit$~MPa anywhere in the line, including the cathode face. This event ends the run. We record the time $t_{\mathrm{nuc}}$, the position $x_{\mathrm{nuc}}$ and the site. A site within one cell of $x = 0$ is a cathode site. Any other site is interior, and it is assigned to a tensile flux divergence if a boundary where $\kappa$ increases in the $+x$ direction lies within one cell of it. Compressive stress is limited by a hillock or extrusion threshold $\sigma_{\mathrm{cc}} = -\PSigCC$~MPa, the signed compressive criterion of Knowlton.\cite{knowltonthesis} Below it, a stiff relief term $+k_r\, r(\sigma_{\mathrm{cc}} - \sigma)$, which raises $\sigma$ back toward $\sigma_{\mathrm{cc}}$, is added to the right side of Eq.~(\ref{eq:stress}), with $r(u) = 0$ for $u \le 0$, $u^{2}/2w$ for $0 < u < w$ and $u - w/2$ for $u \ge w$. The rate is $k_r = \PReliefRate\,\max\kappa/\Delta x^{2}$ and the width is $w = \PReliefWidth$~MPa. Both are numerical choices that hold $\sigma$ at $\sigma_{\mathrm{cc}}$, not measured quantities. The first time the threshold is reached is recorded as $t_{\mathrm{hill}}$. The Cu incubation time $t_{\mathrm{inc}}$ is the first time the Cu fraction in the cathode cell falls below $0.1\,c_0$.

\subsection{Random grain-segment maps}

Each line is a chain of grains laid from the cathode to the anode, with lengths $d$ drawn from a lognormal distribution with median $D_{50} = \PDfifty$~\um{}\cite{atakov1993} and $\sigma_{\ln d} = \PSigLn$.\cite{knowltonthesis} A grain with $d \ge W$ spans the width and is a bamboo segment. A grain with $d < W$ is a grain-boundary (polygranular) segment, and its $\kappa$ and $D_s$ are multiplied by a factor drawn from a lognormal distribution with median 1 and $\sigma_{\ln} = \PSigKappa$, the spread of $\ln d$, because the grain-boundary density scales as $1/d$. The spread of grain-boundary diffusivity with misorientation is not included, so this is a lower bound. We do not impose a cluster-length distribution. Polygranular clusters are runs of grains with $d < W$, and their lengths follow from the grain statistics, as in Knowlton's maps, which were cut from a lognormal film.\cite{knowltonthesis,joo1994}

\begin{figure}[t]
\includegraphics[width=\columnwidth]{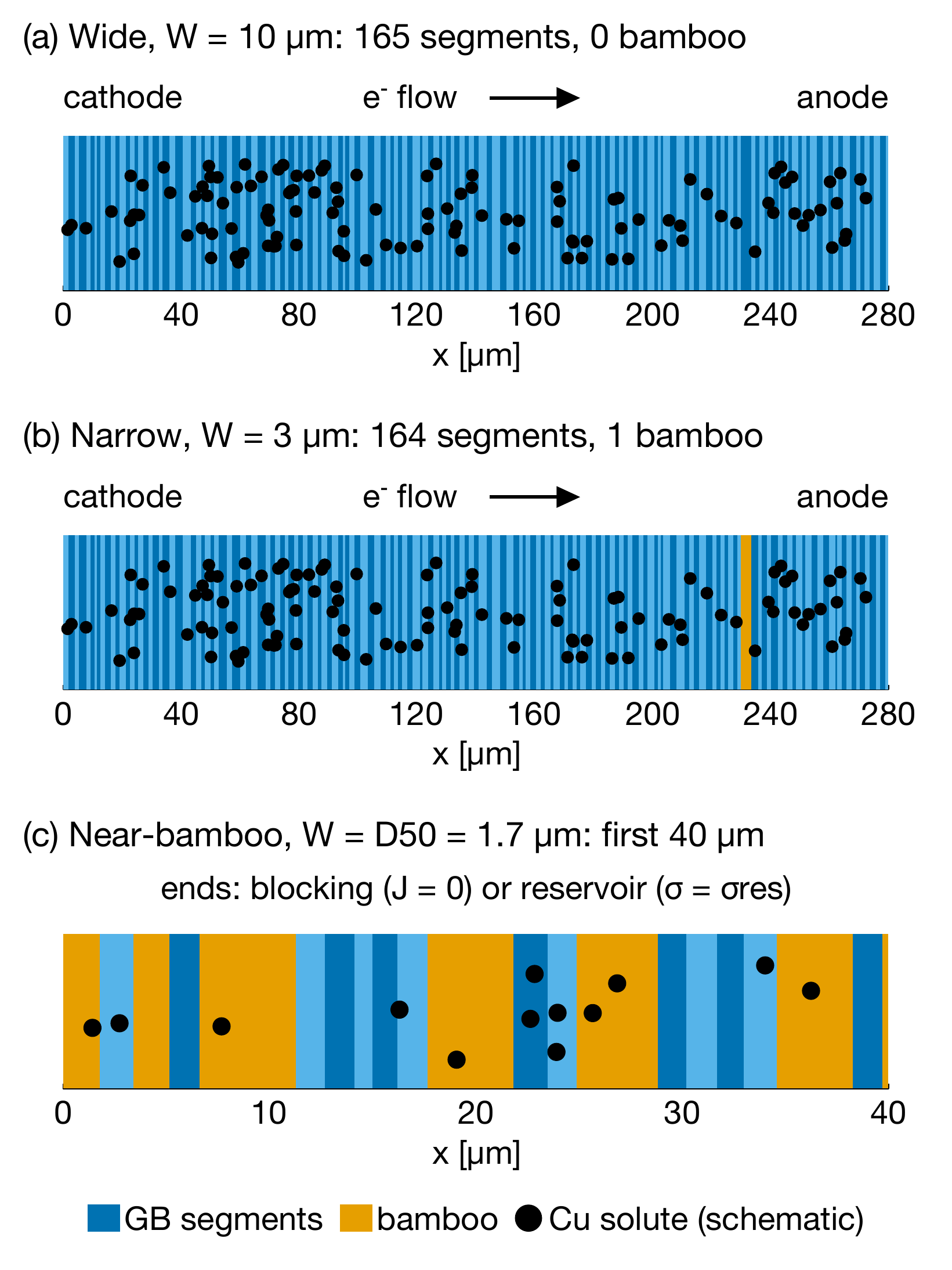}
\caption{Model geometry. Seed-0 grain-segment maps for (a) a wide line ($W = 10$~\um, 165 segments, all grain-boundary segments) and (b) a narrow line ($W = 3$~\um, 164 segments, one bamboo segment), with $L = 280$~\um, and (c) the first 40~\um{} of a near-bamboo line ($W = D_{50}$). Blue bands are grain-boundary segments (alternating shades mark neighboring grains) and orange bands are bamboo segments. Black dots stand for the Cu solute and are schematic only; the model carries a continuous Cu fraction $c(x,t)$. Electrons flow from the cathode ($x = 0$) to the anode ($x = L$). Each end is either blocking (zero atomic flux) or a reservoir (stress held at the residual stress).}
\label{fig:schematic}
\end{figure}

Three width classes are simulated at $L = 280$~\um{} and 200~$^{\circ}$C, the conditions of the experiments (Sec.~\ref{sec:experiment}). The \emph{wide} class has $W = \PWideW$~\um{} ($W/D_{50} = \PWDWide$); bamboo grains essentially never occur, so the line is polygranular, and its flux divergences are the boundaries between grains with different multipliers. The \emph{narrow} class has $W = \PNarrowW$~\um{} ($W/D_{50} = \PWDNarrow$); the probability that a grain spans the width is only \PPBambooNarrow, so the line is mostly polygranular with sparse blocking grains. The \emph{near-bamboo} class has $W = D_{50}$, within the range simulated by Knowlton;\cite{knowltonthesis} half the grains are bamboo. A \emph{uniform} line, one grain-boundary segment with multiplier 1, serves as the deterministic reference. Figure~\ref{fig:schematic} shows examples.

\begin{table*}
\caption{Model inputs. Values at 200~$^{\circ}$C are quoted where the input is a prefactor. ``Derived'' marks a value computed from the cited source; the derivation is recorded with the value in params.yaml of the code release. The thesis of Knowlton\cite{knowltonthesis} is a secondary source for several entries and cites the primary measurements.}
\label{tab:params}
\begin{ruledtabular}
\begin{tabular}{llp{4.3cm}p{5.2cm}}
Symbol & Quantity & Value & Source \\
\hline
$E_{a,\mathrm{gb}}$ & Al grain-boundary activation energy & \PEaGB~eV & Ref.~\onlinecite{knowltonthesis}, p.~128 \\
$D_{0,\mathrm{gb}}$ & Al grain-boundary prefactor (pure Al) & \PDzeroGB~m$^{2}$/s & derived, Ref.~\onlinecite{knowltonthesis}, pp.~74, 76 \\
$E_{a,b}$ & Bamboo (interface) activation energy & \PEaB~eV & Ref.~\onlinecite{knowltonthesis}, p.~128 \\
$D_{0,b}$ & Bamboo prefactor & \PDzeroB~m$^{2}$/s & derived, Ref.~\onlinecite{knowltonthesis}, pp.~74, 76 \\
$\Zs_{\mathrm{Al}}$ & Al effective charge (magnitude) & \PZAl & Refs.~\onlinecite{knowltonthesis,korhonen1993,clement1995} \\
$|\Zs_{\mathrm{Cu}}|$ & Cu effective charge in Al (magnitude) & \PZCu & Ref.~\onlinecite{kao2001} \\
$E_{a,\mathrm{Cu}}$ & Cu grain-boundary activation energy & \PEaCu~eV & Ref.~\onlinecite{kao2001} \\
$D_{0,\mathrm{Cu}}$ & Cu grain-boundary prefactor & \PDzeroCuSI~m$^{2}$/s & Ref.~\onlinecite{kao2001} (factor of 40 uncertain) \\
$\beta$ & Cu blocking constant, Eq.~(\ref{eq:block}) & \PBeta & derived, Ref.~\onlinecite{knowltonthesis}; form, Refs.~\onlinecite{rosenberg1972,park1999} \\
$c_{\mathrm{sol}}$ & Cu solubility in Al at 200~$^{\circ}$C & \PCsolWt~wt\% & derived, Ref.~\onlinecite{knowltonthesis}, pp.~42, 137; Ref.~\onlinecite{murray1985} \\
$\rho_0$ & Resistivity at $T_0 = \PRhoTzero\,^{\circ}$C & \PRhoZero~$\mu\Omega$\,cm & Ref.~\onlinecite{knowltonthesis}, p.~76 \\
$\alpha_\rho$ & Temperature coefficient of $\rho$ & \PAlphaRho~K$^{-1}$ & derived, Ref.~\onlinecite{knowltonthesis}, p.~70 \\
$B$ & Effective modulus & \PB~GPa & Ref.~\onlinecite{knowltonthesis}, p.~76 \\
$\Omega$ & Atomic volume & \POmega~m$^{3}$ & Ref.~\onlinecite{knowltonthesis}, p.~76 \\
$\sigma_{\mathrm{crit}}$ & Void-nucleation threshold (tensile) & \PSigCrit~MPa & Ref.~\onlinecite{knowltonthesis} (quoted range 100~MPa to 1~GPa) \\
$\sigma_{\mathrm{cc}}$ & Hillock threshold (compressive) & $-$\PSigCC~MPa & Ref.~\onlinecite{knowltonthesis}, pp.~127--129 \\
$D_{50}$ & Median grain size & \PDfifty~\um & Ref.~\onlinecite{atakov1993}, as reported in Ref.~\onlinecite{knowltonthesis} \\
$\sigma_{\ln d}$ & Lognormal spread of grain size & \PSigLn & Ref.~\onlinecite{knowltonthesis}, p.~94 \\
$\sigma_{\ln}$ & Spread of the segment $\kappa$ multiplier & \PSigKappa & derived from $\sigma_{\ln d}$ (assumption) \\
$W$, $L$, $T$ & Width, length, temperature & \PWideW{} and \PNarrowW~\um, 280~\um, 200~$^{\circ}$C & Ref.~\onlinecite{dingari2007} \\
$W_{\mathrm{nb}}$ & Near-bamboo width & $\PWDNB\,D_{50}$ & within the range of Ref.~\onlinecite{knowltonthesis} \\
$k_r$, $w$ & Relief rate factor and width & \PReliefRate, \PReliefWidth~MPa & numerical assumption \\
\hline
\multicolumn{4}{l}{Cu(Mn) lines (Sec.~\ref{sec:res:cumn})} \\
$E_a$, $D_0$ & Cu interface diffusion & \PEaCuHost~eV, \PDzeroCuHost~m$^{2}$/s & Ref.~\onlinecite{alam2005}, as tabulated in Ref.~\onlinecite{alshohel2021} \\
$\Zs$, $\rho$, $B$, $\Omega$ & Cu host & \PZCuHost; \PRhoCu~$\Omega$\,m; \PBCu~GPa; \POmegaCu~m$^{3}$ & Ref.~\onlinecite{alam2005}, as tabulated in Ref.~\onlinecite{alshohel2021} \\
$\sigma_{\mathrm{crit}}$ & Void-nucleation threshold in Cu & \PSigCritCu~MPa & Ref.~\onlinecite{alam2005}, as tabulated in Ref.~\onlinecite{alshohel2021} \\
$D_{0}$, $E_a$ & Mn lattice diffusion in Cu & \PDzeroMnSI~m$^{2}$/s, \PEaMn~eV & Ref.~\onlinecite{butrymowicz1975} \\
$\beta_{\mathrm{Mn}}$ & Mn blocking constant & \PBetaMn & derived from Ref.~\onlinecite{hu2012} (range \PBetaMnLow{} to \PBetaMnHigh) \\
\end{tabular}
\end{ruledtabular}
\end{table*}

%% file: sections/methods.tex
\section{Numerical method and validation}
\label{sec:methods}

\subsection{Discretization}

Equations~(\ref{eq:stress}) and (\ref{eq:solute}) are solved by the method of lines on a uniform cell-centered finite-volume grid of $N$ cells of width $\Delta x = L/N$; all production runs use $N = 1600$ ($\Delta x = 0.175$~\um). Each cell takes the length-weighted harmonic mean of the $\kappa$ (and $D_s$) of the segments that overlap it, and it counts as grain-boundary material, to which Eq.~(\ref{eq:block}) applies, when at least half of it is grain boundary. The stress flux on an interior face $f$ between cells $f-1$ and $f$ is
\begin{equation}
F_f = \kappa_f\left(\frac{\sigma_f - \sigma_{f-1}}{\Delta x} + G\right),\qquad
\kappa_f = \frac{2\kappa_{f-1}\kappa_f}{\kappa_{f-1} + \kappa_f},
\label{eq:fv}
\end{equation}
where the cell values $\kappa_i$ already include the Cu blocking of Eq.~(\ref{eq:block}) evaluated with the local $c_i$. The harmonic face value keeps the flux continuous across the contrast of up to $10^{4}$ between grain-boundary and bamboo cells. A blocking end has $F = 0$; a reservoir end uses a half-cell face, $F_0 = \kappa_0[(\sigma_0 - \sigma_{\mathrm{res}})/(\Delta x/2) + G]$ and likewise at $x = L$. Then $\mathrm{d}\sigma_i/\mathrm{d}t = (F_{i+1} - F_i)/\Delta x$, plus the relief sink.

The Cu flux uses the Scharfetter-Gummel form,\cite{scharfetter1969} which solves the drift-diffusion flux exactly on each face for constant flux:
\begin{equation}
\begin{gathered}
J_f = \frac{D_{s,f}}{\Delta x}\left[\mathcal{B}(-v\Delta x)\, c_{m,f-1} - \mathcal{B}(v\Delta x)\, c_{m,f}\right],\\
\mathcal{B}(z) = \frac{z}{e^{z} - 1},
\end{gathered}
\label{eq:sg}
\end{equation}
with harmonic face diffusivities $D_{s,f}$, the mobile Cu $c_m$ of the solubility cap in each cell, and $J = 0$ on both end faces. The scheme reduces to central diffusion as $v \to 0$ and to upwinding for $|v\Delta x| \gg 1$, keeps $c \ge 0$, conserves the Cu inventory exactly, and has the exact discrete steady state $c_i \propto e^{v x_i}$ on any grid.

\subsection{Time integration and stopping}

The $2N$ ordinary differential equations are scaled ($x/L$, $t\,\kappa_{\mathrm{max}}/L^{2}$, $\sigma/GL$, $c/c_0$) and integrated with the variable-order backward differentiation formula (BDF) of SciPy's \texttt{solve\_ivp},\cite{virtanen2020} with relative and absolute tolerances of $10^{-6}$ and $10^{-9}$ and an analytic sparse Jacobian. The Jacobian has tridiagonal stress-stress, stress-Cu and Cu-Cu blocks and a zero Cu-stress block. Void nucleation, the hillock threshold and Cu incubation are root-found events of the integrator. The scheduled horizon is $t_{\mathrm{end}} = 5\max(L^{2}/\kappa_{\mathrm{min}}, L^{2}/D_{s,\mathrm{min}})$, five times the slowest diffusion time of the line. Where the Cu content exceeds the solubility, Cu redistribution is slower by about $c_0/c_{\mathrm{sol}}$, and the horizon is lengthened by that factor. The run is integrated in 40 logarithmically spaced chunks between $10^{-6}t_{\mathrm{end}}$ and $t_{\mathrm{end}}$. After each chunk the current rates are extrapolated linearly over the remaining horizon; the run stops as steady when, at two consecutive checks, $c$ would change by less than $10^{-3}c_0$ and the stress would close less than one tenth of the remaining margin to $\sigma_{\mathrm{crit}}$. Cells held by the relief term are excluded from the stress rate. Because the rates decay toward the steady state, the linear extrapolation is conservative. A steady-stopped line cannot nucleate before $t_{\mathrm{end}}$ and is recorded as immortal, right-censored at $t_{\mathrm{end}}$ (Sec.~\ref{sec:stats}).

\subsection{Validation}
\label{sec:validation}

\begin{figure}[t]
\includegraphics[width=\columnwidth]{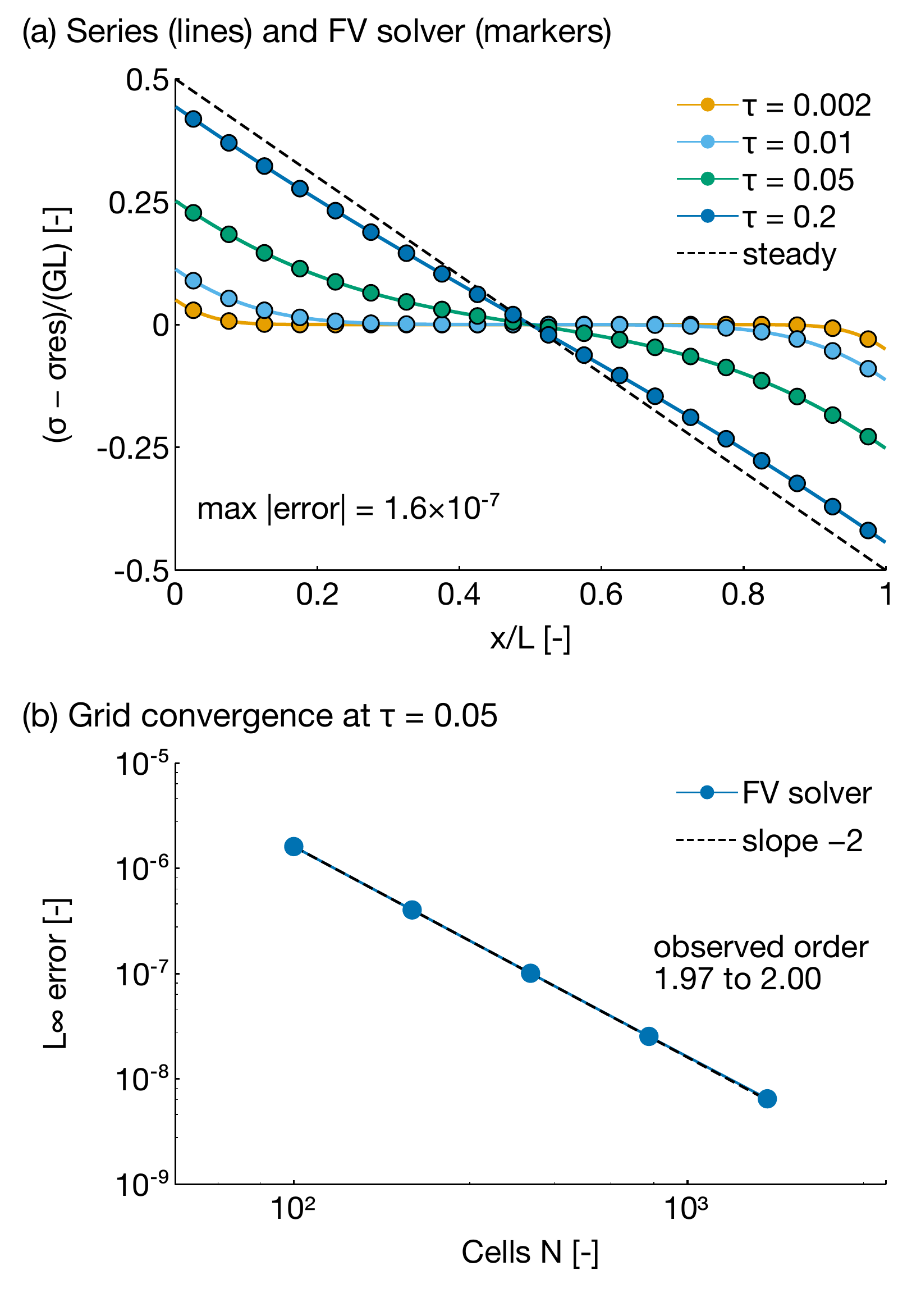}
\caption{Verification against the Korhonen series for a uniform line with blocking ends and constant $G$. (a) Normalized stress $(\sigma - \sigma_{\mathrm{res}})/GL$ against $x/L$ at dimensionless times $\tau = \kappa t/L^{2}$ of 0.002, 0.01, 0.05 and 0.2: the analytic series (lines) and the finite-volume solver with \NVoneCells{} cells (markers, every 80th cell shown), with the steady state $1/2 - x/L$ (dashed). The largest difference between solver and series over all cells and all four times is \NVoneMaxErr{} of $GL$ (at $\tau = \NVoneMaxErrTau$). (b) Maximum error of the solver against the series at $\tau = 0.05$ as a function of the number of cells $N$, with a slope $-2$ reference line (dashed). The observed order is \NConvOrderMin{} to \NConvOrderMax.}
\label{fig:validation}
\end{figure}

\emph{V1, analytic solution.} For a uniform line with blocking ends and constant $G$, the normalized stress $\hat\sigma = (\sigma - \sigma_{\mathrm{res}})/GL$ obeys\cite{korhonen1993}
\begin{equation}
\hat\sigma(\xi,\tau) = \frac{1}{2} - \xi - \sum_{n\ \mathrm{odd}} \frac{4}{n^{2}\pi^{2}}\cos(n\pi\xi)\, e^{-n^{2}\pi^{2}\tau},
\label{eq:series}
\end{equation}
with $\xi = x/L$ and $\tau = \kappa t/L^{2}$. Figure~\ref{fig:validation}(a) compares the solver ($N = \NVoneCells$) with Eq.~(\ref{eq:series}) at four times; the largest difference over all cells and times is \NVoneMaxErr{} of $GL$, at $\tau = \NVoneMaxErrTau$. At $\tau = 0.05$ the maximum error falls from \NConvErrCoarse{} at $N = \NConvNmin$ to \NConvErrFine{} at $N = \NConvNmax$, an observed order of \NConvOrderMin{} to \NConvOrderMax{} [Fig.~\ref{fig:validation}(b)]. Tightening the integrator tolerances from $10^{-6}$ to $10^{-9}$ changes the profiles by less than $10^{-4}$ of $GL/2$, and with blocking ends the mean stress is conserved to $10^{-8}GL$ in a line with a diffusivity contrast of $10^{5}$.

\emph{V2, steady state and Blech threshold.} The solver reaches the linear steady state $\hat\sigma = 1/2 - \xi$. In uniform and two-segment test lines the cathode stress reaches $\sigma_{\mathrm{crit}}$ when $GL/2$ exceeds $\sigma_{\mathrm{crit}} - \sigma_{\mathrm{res}}$ by 2\% and does not when it falls 2\% short, and the nucleation time agrees with the root of Eq.~(\ref{eq:series}) to 1\%. The reservoir boundary keeps a uniform line flat, and a three-segment line with reservoir ends reaches its analytic piecewise-linear steady state. The Cu scheme conserves mass, stays non-negative and reaches the exponential steady state of Eq.~(\ref{eq:solute}).

\emph{Grid convergence on random maps.} The same \NGCMaps{} maps per condition were run with $N = 800$, 1600 and 3200 cells, at 0.5~wt\% Cu, for every class and end condition at 3~\MAcm{} and for wide lines with reservoir ends at 2~\MAcm. Between $N = 1600$ and 3200 the median change of $t_{\mathrm{nuc}}$ per map is \NGCMedMin\% to \NGCMedMax\%, and the median lifetime of a condition changes by at most \NGCTfiftyMax\%. \NGCFlip{} of the \NGCRuns{} runs change between nucleating and immortal. In \NGCSiteMin\% to \NGCSiteMax\% of the narrow and near-bamboo runs the nucleation site moves by more than 1~\um, to a competing divergence of nearly equal stress, so the site distributions should be read as statistics over such competing sites. The same set at 2~wt\% Cu, where the Cu steps are largest, changes the median lifetime of a condition by at most \NGridTwoMaxT\% and moves \NGridTwoClass{} runs between nucleating and immortal.

\emph{Steady stop.} Once the Cu is capped, nearly all the runs that stop steady are Cu-free: \NEthreeSteady{} E3 runs, none of them with Cu, and a few E9 runs with Cu. Could the stop hide a late failure? We took \NSteadyRuns{} steady-stopped runs (every E9 run with Cu, plus a stratified sample of the Cu-free ones) and ran them again to the full horizon with the stop switched off. Not one of them nucleated. The hillock times came out the same, and the minimum stress moved by \NSteadyDsig~MPa at most. The number of full-horizon runs that failed numerically was \NSteadyFailed.

\emph{Regression guard.} A stratified sample of \NRegSample{} runs from the committed results is re-simulated by the test suite, which requires nucleation and hillock times to agree to a relative $10^{-6}$ and the sites and immortality to be identical.

\subsection{Statistics}
\label{sec:stats}

Lines that do not nucleate are right-censored at their scheduled horizon (Type-I censoring). We call such lines immortal: no void nucleates within the horizon, or the steady stop shows that the stress has stopped rising below $\sigma_{\mathrm{crit}}$. This is immortality within the model, not a demonstrated infinite lifetime. Per condition we fit a lognormal distribution to the nucleation times by maximum likelihood with censoring, giving the median $t_{50}$ and the spread $\sigma_{\ln}$, and we compute the Kaplan-Meier median\cite{kaplan1958} with a 95\% confidence interval from the linear Greenwood band. Where the immortal fraction is \NImmGate{} or more, a single lognormal is mis-specified, because the immortal lines form a separate sub-population, and the Kaplan-Meier median is reported instead; where the immortal fraction is 0.5 or more, the median itself is infinite. The early-failure time $t_{01}$ is the 1\% quantile of the population; its 95\% interval, and that of $t_{01}$ divided by the median, are percentile bootstrap intervals over the maps (2000 resamples).

Black's-law parameters are fitted on the per-run lifetimes with a censored lognormal accelerated-failure-time model,
\begin{equation}
\ln t = \ln A - n \ln j + \frac{E_a}{\kB T} + \sigma\varepsilon,\qquad \varepsilon \sim \mathcal{N}(0,1),
\label{eq:aft}
\end{equation}
by maximum likelihood, with 95\% Wald intervals from the observed information. Because a map with a given seed is used at every current density, the runs of one map are correlated; a bootstrap that resamples whole maps (\NBootB{} resamples) gives intervals \NBootWidthMin{} to \NBootWidthMax{} times the Wald width, and no conclusion about $n$ changes. $E_a$ enters only when $T$ varies. We fit three current-density ranges separately: 1 to 3, 2 to 5 and 4 to 6~\MAcm. In a range where the immortal fraction reaches \NImmGate{} at any current density we give no $n$, since there the slope tracks how fast immortality grows and is not a Black exponent. An ordinary least-squares fit of $\ln t_{50}$ serves as a cross-check.

\subsection{Simulations}

All runs use Al-Cu parameters at 200~$^{\circ}$C and $L = 280$~\um{} unless otherwise stated. E1 is the uniform line with blocking ends at $j = 1$ to 6~\MAcm, $T = 150$ to 300~$^{\circ}$C, Cu contents of 0 and 0.5~wt\% and $\sigma_{\mathrm{res}} = 0$, 37.5 and 75~MPa (\NEoneRuns{} runs). E2 follows the stress and Cu fields of \NEtwoMaps{} maps per class, end type and Cu content (0, 0.5, 1 and 2~wt\%) at 3~\MAcm. E3 is the population study: three classes, Cu contents of 0, 0.5 and 2~wt\%, $j = 1$ to 6~\MAcm{} and both end types, with \NEthreeMaps{} maps per condition (\NEthreeRuns{} runs; within a class, the map with a given seed is the same in every condition). A sensitivity set without compressive relief repeats the wide and narrow classes at 0.5~wt\% Cu with \NNoReliefMaps{} maps (\NNoReliefRuns{} runs). E4 varies $L$ from 1 to 50~\um{} at 2~\MAcm{} and 0.5~wt\% Cu with blocking ends (\NEfourMaps{} maps per class and length). It treats all Cu as mobile: with the cap, the precipitate of a short line held at steady state piles up in the anode cells (to about 100 times $c_0$ in a few cells, within the finite Cu inventory) and makes the integration orders of magnitude slower, while the Blech threshold depends on $jL$ alone (Sec.~\ref{sec:res:cutransport}). A check with the cap repeats $L = 3$ to 6~\um{} for wide and narrow lines on \NEfourCapMaps{} maps (\NEfourCapRuns{} runs). E5 is the Black's-law analysis of the E3 populations (Sec.~\ref{sec:res:e5}). E6 applies the solver to Cu(Mn) lines (Sec.~\ref{sec:res:cumn}). E7 and E3\textsubscript{mobile} are the limiting cases of Cu transport (Sec.~\ref{sec:res:cutransport}). E7 repeats the E3 conditions with 0.5 and 2~wt\% Cu on the first \NEsevenMaps{} maps of each class with the Cu held immobile at $c_0$ (\NEsevenRuns{} runs). E3\textsubscript{mobile} repeats the E3 runs with Cu with all Cu treated as mobile solute, without the cap (\NEthreeMobRuns{} runs). E10 repeats the E7 grid with the cap and blocking either by the total Cu (the main model, as a check against E3) or by the dissolved Cu only (\NTenRuns{} runs). E11 repeats the E10 grid with blocking by the dissolved Cu plus a fraction $f = 0.25$, 0.5 or 0.75 of the precipitate (\NElevenRuns{} runs). E12 continues the Cu field alone to the durations of the measured tests (Sec.~\ref{sec:comparison}); the Cu equation has no stress term, so this needs no assumption about the stress after nucleation (\NTwelveMaps{} maps per case, at 200, 225 and 250~$^{\circ}$C, and with all Cu mobile at 200~$^{\circ}$C). E8 changes one input at a time for the uniform line with 0.5~wt\% Cu and blocking ends at 2 and 5~\MAcm{} (\NEightRuns{} runs; Sec.~\ref{sec:discussion}). E9 repeats the E3 conditions at 150, 250 and 300~$^{\circ}$C on the first \NEnineMaps{} maps of each class (\NEnineRuns{} runs; Sec.~\ref{sec:res:e9}). In all, \NTotalRuns{} simulations were run, all seeded and reproducible, and none failed numerically. The full grid takes about 12~h on 12 cores of a laptop.

\subsection{Use of AI tools}
\label{sec:ai}
Claude Code version 2.1.316 (Anthropic), an AI coding agent running the large language model Claude Opus 5.5 (Anthropic; model ID claude-opus-5-5), was used to write and test the simulation and analysis code, to propose simulation and statistical designs based on the corresponding author's earlier experimental work (censored runs, Kaplan-Meier medians, accelerated-failure-time fits) and to implement the designs the corresponding author adopted, to run the simulations, to search and summarize the literature, and to draft the manuscript text. It was used to shorten the time needed to build and run a solver and analysis pipeline of this size and to make every number and figure regenerable from scripts. The outputs were checked by more than 200 automated tests, by the analytic validation of Sec.~\ref{sec:validation}, by a regression guard that re-simulates a sample of the committed results, and by code review in separate Claude Code sessions (Claude Opus 5.5 or Claude Sonnet 5.5, Anthropic) that had not written the code under review. Every citation was resolved through its DOI or a resolvable URL. All simulation data come from the documented, seeded code; AI did not generate or alter data or images in any other way. The corresponding author designed the study, including the simulation experiments and the statistical methods, directed the work, and made and checked every scientific decision, including the reading of the results. Both authors checked the results and the text and take responsibility for them.

%% file: sections/experiment.tex
\section{Experiment}
\label{sec:experiment}

These measurements were made for the corresponding author's dissertation,\cite{dingari2007} which first reported the damage and Cu profiles. We use them again here to test the model. The test structures were made by ON Semiconductor. A 0.85~\um{} thick Al-Cu film on a 0.2~\um{} TiW underlayer was patterned into lines 280~\um{} long and either 10~\um{} wide, from Al-2~wt\%~Cu, or 3~\um{} wide, from Al-0.5~wt\%~Cu, and covered by a 1~\um{} SiN passivation layer. The chip had a single metal level. Each end of the test line joined wider terminal lines on the same level, which carried the current and the voltage sense lines, and each chip was wire-bonded into a 24-pin dual in-line package.\cite{dingari2007} The line ends were therefore not W studs or vias. Five wide lines (w1 to w5) were stressed at 200~$^{\circ}$C and 2~\MAcm, and six narrow lines (n1 to n6) at 200~$^{\circ}$C and 5~\MAcm. The resistance was recorded during stressing, and the line temperature was obtained from the resistance through its temperature coefficient, which corrects for Joule heating. Each line was stressed to failure as judged from its resistance; the wide lines failed after about 6~days and the narrow lines after about 11~days. One further wide line, measured by EDS before and after stressing, failed as an open circuit after 120~h. As deposited, the Cu was not uniform along the line, but after an anneal at 200~$^{\circ}$C for 24~h without current it was uniform along two wide lines measured by EDS.

After stressing we located the damage on each line by optical microscopy, atomic force microscopy and scanning electron microscopy (SEM). The Cu was measured by energy-dispersive x-ray spectroscopy (EDS) in the SEM at 18 to 20~keV. On each line we chose 6 to 10 areas spaced along its length and rastered the beam over each one for 20 to 30~min. The dissertation gives the area as roughly 10~\um{} $\times$ 25~\um{} and shows it on a 10~\um{} line;\cite{dingari2007} it gives no separate size for the 3~\um{} lines, where the 10 areas were centered about 30~\um{} apart. No area reached the neighboring line. If an area is wider than the line, only the part over the line gives Cu counts. That lowers the signal by about the same factor at every position and leaves the shape of the relative profile unchanged. We removed the background and fitted the Cu K$\alpha$ peak together with the nearby W L$\alpha$ peak of the underlayer, taking the Cu peak height as proportional to the mean Cu concentration of the area. The profiles are therefore relative. They were recorded against distance from the anode; below we convert this to $x/L$ measured from the cathode.

%% file: sections/results.tex
\section{Results}
\label{sec:results}

\subsection{Uniform line (E1)}
\label{sec:res:e1}

For the uniform line with blocking ends, the fitted current-density exponent is $n = \NEoneNMin$ to \NEoneNMax{} over all current-density ranges, residual stresses and both Cu contents (each fit is made over all temperatures at once, with $n$ and $E_a$ fitted together). Without Cu the apparent activation energy is \NEoneEaCuZeroMin{} to \NEoneEaCuZeroMax~eV, above the input $E_{a,\mathrm{gb}} = \PEaGB$~eV because $\kappa$ contains $1/T$ and $G^{2}$ contains $\rho(T)^{2}$. These values follow from the early-time form of Eq.~(\ref{eq:series}). At the cathode $\hat\sigma(0,\tau) \approx 2\sqrt{\tau/\pi}$, so for $\sigma_{\mathrm{crit}} \ll GL/2$ the nucleation time is $t_{\mathrm{nuc}} \approx \pi\sigma_{\mathrm{crit}}^{2}/(4G^{2}\kappa)$, which scales as $j^{-2}$ and does not depend on $L$. At 3~\MAcm{} and 200~$^{\circ}$C, $GL = \PGLThree$~GPa, and the estimate gives \PTnucEstThree~h, equal to the simulated \NEoneTCuZeroSZero~h. Nucleation therefore probes only a cathode layer of order $\sigma_{\mathrm{crit}}/G = \PEllThree$~\um{} at 3~\MAcm, about one grain.

With 0.5~wt\% Cu the nucleation time at 3~\MAcm{} rises from \NEoneTCuZeroSZero{} to \NEoneTCuHalfSZero~h, a factor of \NEoneCuFactor. This is somewhat less than the uniform blocking factor $1 + \beta c_0 = \PBlockHalf$: part of the Cu leaves the cathode layer as dissolved Cu while the stress builds up. The immobile-Cu control of Sec.~\ref{sec:res:cutransport} confirms this. The apparent activation energy rises to \NEoneEaCuHalfMin{} to \NEoneEaCuHalfMax~eV, from \NEoneEaCuHalfSresZero~eV at $\sigma_{\mathrm{res}} = 0$ to \NEoneEaCuHalfSresMax~eV at \PSresMax~MPa. Two effects raise it: Cu transport has the higher activation energy (\PEaCu~eV against \PEaGB~eV), and the dissolved fraction grows with temperature along the solvus (\PCsolWtOneFifty{} to \PCsolWtThreeHundred~wt\% from 150 to 300~$^{\circ}$C), so at higher $T$ relatively more Cu can leave the cathode layer before nucleation. The exponent $n$ is unchanged by Cu in this line.

\subsection{Stress and Cu fields (E2)}
\label{sec:res:e2}

\begin{figure}[t]
\includegraphics[width=\columnwidth]{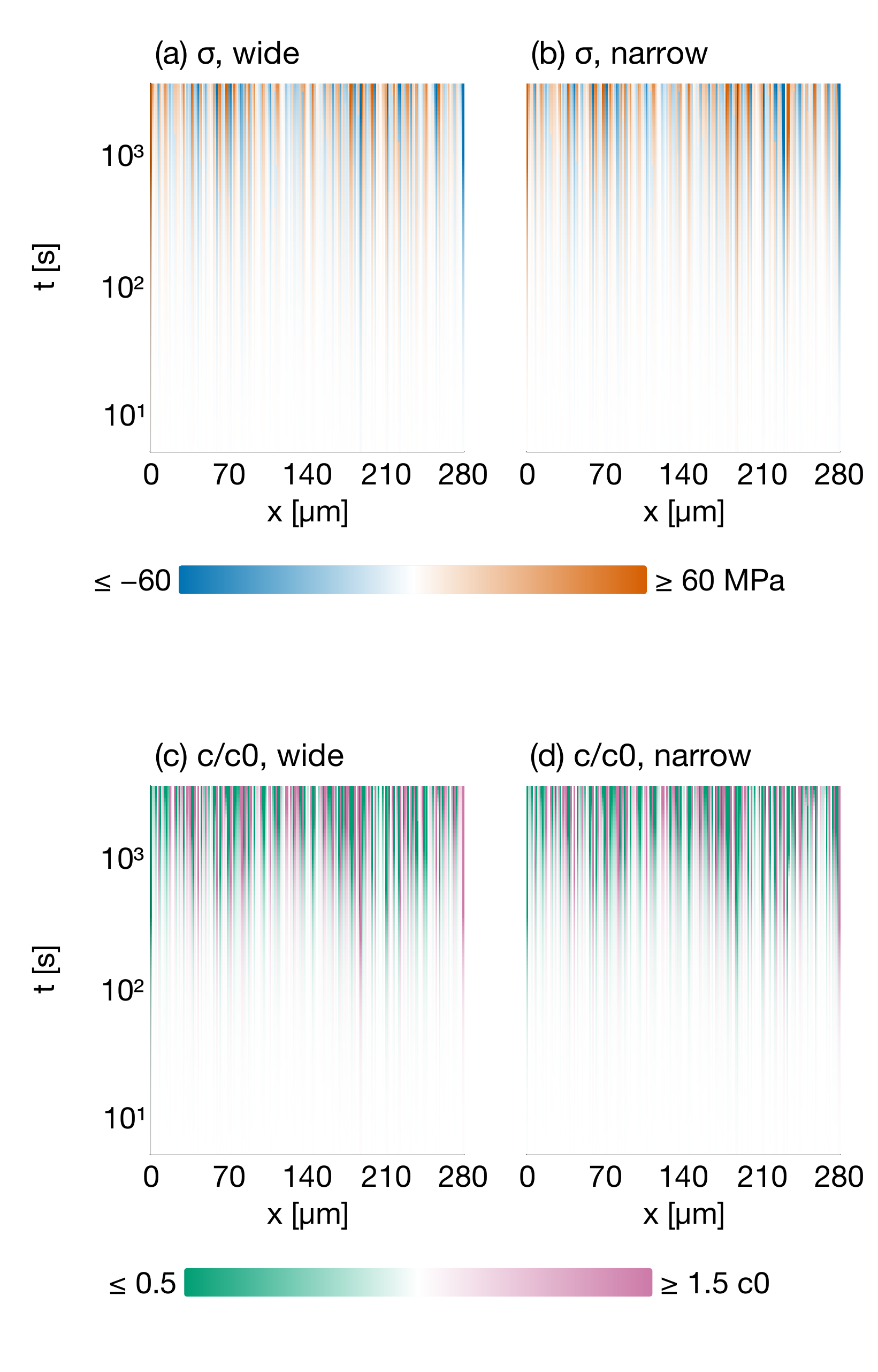}
\caption{Stress and Cu evolution in single lines, Al-0.5~wt\% Cu, blocking ends, $j = 3$~\MAcm, seed-0 maps. (a, b) Hydrostatic stress $\sigma(x,t)$ in MPa and (c, d) total Cu concentration $c(x,t)/c_0$ (dissolved and precipitated) for (a, c) the wide line and (b, d) the narrow line, on a logarithmic time axis from $t = 5$~s to void nucleation at the cathode ($t_{\mathrm{nuc}} = \NSeedTnucWide$~s for both lines). The color scales are clipped at $\pm 60$~MPa and at 0.5 to 1.5~$c_0$; the cathode stress reaches $\sigma_{\mathrm{crit}} = \PSigCrit$~MPa at $t_{\mathrm{nuc}}$, and the largest interior Cu content is \NSeedPeakWide~$c_0$. Each column of the map pools 8 cells (1.4~\um), keeping the value farthest from $\sigma = 0$ or $c = c_0$ so that narrow peaks at flux-divergence sites remain visible.}
\label{fig:fields}
\end{figure}

Figure~\ref{fig:fields} follows the seed-0 wide and narrow lines with 0.5~wt\% Cu and blocking ends at 3~\MAcm. Both nucleate at the cathode, at \NSeedTnucWide~s. When the stress reaches $\sigma_{\mathrm{crit}}$ it is still confined to the cathode end and a few flux divergences. The Cu field has changed along the whole line by then. Only the dissolved part can move, $c_{\mathrm{sol}}/c_0 = \PMobFracHalf$ of the Cu at 0.5~wt\%. But where the dissolved flux diverges, the precipitate dissolves or grows, and the total Cu picks up sharp local dips and peaks at the boundaries where $\kappa$ changes. Precipitate does not diffuse back, so nothing smooths these features out. In the seed-0 maps with blocking ends the largest interior Cu content at nucleation is \NSeedPeakWide~$c_0$ (at $x/L = \NSeedPeakXWide$), and across all seed-0 E2 runs with Cu the largest local content is \NCuMaxNucMin{} to \NCuMaxNucMax~$c_0$, at most \NCuMaxNucPct~at.\%. Across all E2 maps with blocking ends and Cu, the Cu fraction in the cathode cell fell below $0.1c_0$ before nucleation in \NTincFinBlk{} of \NEtwoBlkRuns{} runs, so the incubation stage often ends shortly before nucleation. With reservoir ends the wide lines live longer, and the median $t_{\mathrm{inc}}$ is \NTincWideRes~h.

With the solubility cap, Cu builds up toward the anode only weakly. The median ratio of the mean Cu fraction in the last tenth of the wide line to $c_0$ is \NAnodeWideBlkCuHalf, \NAnodeWideBlkCuOne{} and \NAnodeWideBlkCuTwo{} for 0.5, 1 and 2~wt\% with blocking ends, and \NAnodeWideResCuHalf, \NAnodeWideResCuOne{} and \NAnodeWideResCuTwo{} with reservoir ends. The drift flux of dissolved Cu is limited to $v c_{\mathrm{sol}}$, so the amount carried to the anode does not grow with the Cu content. In the near-bamboo class the ratio is \NAnodeNBBlkCuHalf, because bamboo grains, where Cu moves slowly, interrupt the drift. The positions of interior Cu peaks are compared with the measurements in Sec.~\ref{sec:comparison}.

\subsection{Lifetime distributions (E3)}
\label{sec:res:e3}

\begin{figure}[t]
\includegraphics[width=\columnwidth]{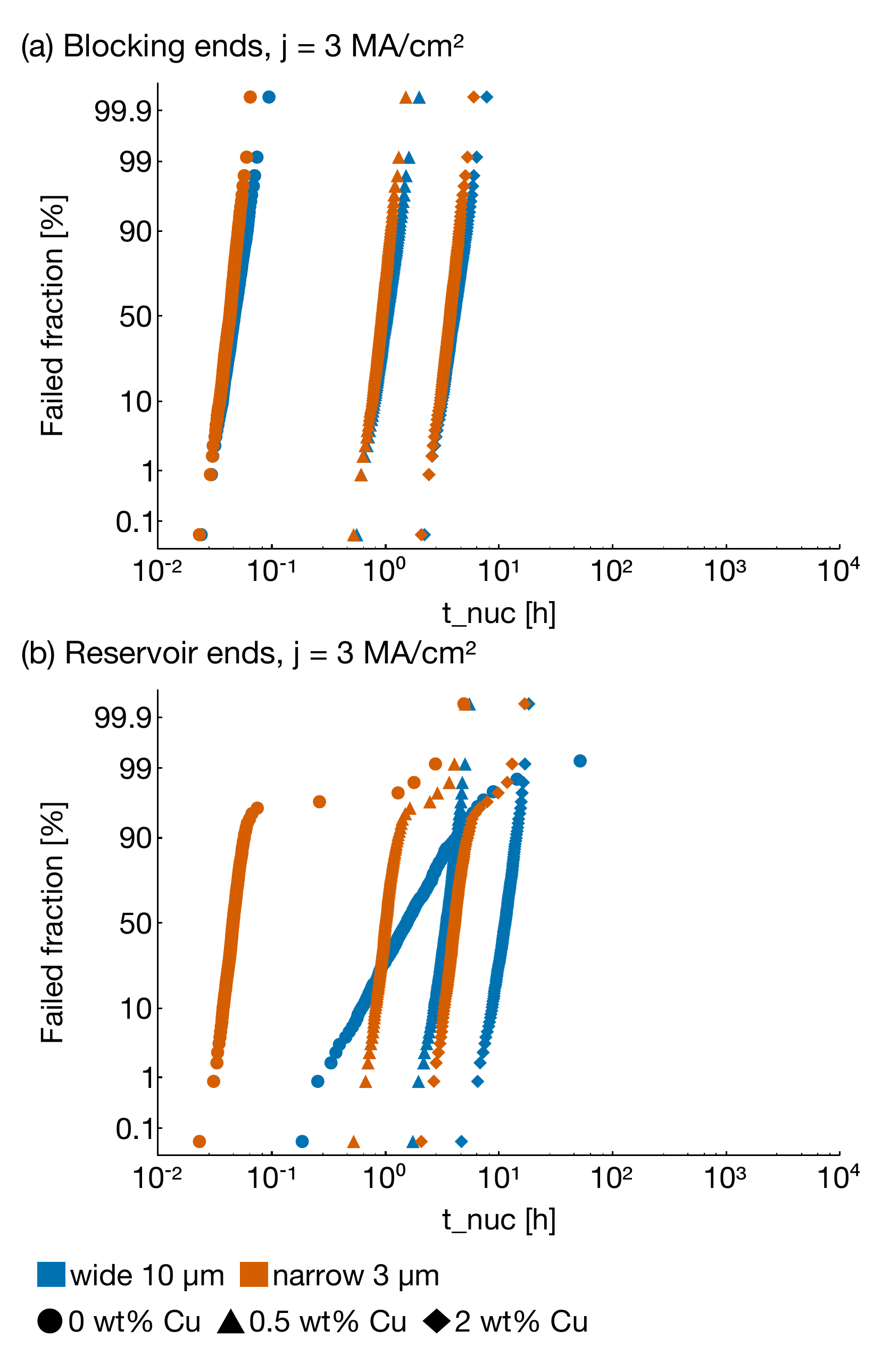}
\caption{Lognormal probability plot of the void-nucleation time $t_{\mathrm{nuc}}$ (h) for wide (blue) and narrow (vermilion) lines with 0, 0.5 and 2~wt\% Cu (symbols), $j = 3$~\MAcm, \NEthreeMaps{} maps per condition, with (a) blocking and (b) reservoir ends. Immortal runs are right-censored at their scheduled horizon, and the plotting positions are Kaplan-Meier midpoints. With reservoir ends the immortal fraction of the Cu-free wide lines is \NImmWideResCuZeroJThree; every line with Cu fails.}
\label{fig:lognormal}
\end{figure}

Table~\ref{tab:t50} lists the median nucleation times at 3~\MAcm, and Fig.~\ref{fig:lognormal} shows the distributions. With blocking ends the distributions are narrow: $\sigma_{\ln} = \NSigBlkWNMin$ to \NSigBlkWNMax{} for wide and narrow lines at every current density, and the 1\% quantile $t_{01}$ is \NTzeroOneRatioWideBlk{} (95\% interval \NTzeroOneRatioWideBlkLo--\NTzeroOneRatioWideBlkHi; wide) and \NTzeroOneRatioNarrowBlk{} (\NTzeroOneRatioNarrowBlkLo--\NTzeroOneRatioNarrowBlkHi; narrow) of the median at 0.5~wt\% Cu and 3~\MAcm. The narrow spread follows from the cathode-layer argument of Sec.~\ref{sec:res:e1}: the lifetime is set by the diffusivity of the first grain or two, whose multiplier varies by $\sigma_{\ln} = \PSigKappa$. Wide and narrow lines then fail at almost the same time. At 0.5~wt\% Cu the ratio of their medians is \NWNRatioBlkMin{} to \NWNRatioBlkMax{} for $j = 2$ to 6~\MAcm.

Copper raises the median steadily with content. At 3~\MAcm{} and blocking ends, 0.5~wt\% multiplies $t_{50}$ by \NCuFacWideBlkHalf{} (wide), \NCuFacNarrowBlkHalf{} (narrow) and \NCuFacNBBlkHalf{} (near-bamboo), and 2~wt\% by \NCuFacWideBlkTwo, \NCuFacNarrowBlkTwo{} and \NCuFacNBBlkTwo, close to the static blocking factors $1 + \beta c_0 = \PBlockHalf$ and \PBlockTwo. With reservoir ends the factor is the same as with blocking ends for the near-bamboo class and close to it for the narrow class, but much smaller for the wide class (\NCuFacWideResHalf{} and \NCuFacWideResTwo), for the reason given next.

With reservoir ends a uniform line stays at $\sigma_{\mathrm{res}}$, so stress develops only from contrasts in $\kappa$. Narrow and near-bamboo lines contain bamboo grains, whose contrast of $10^{4}$ creates strong flux divergences; their reservoir-end lifetimes are close to the blocking-end values, and at most \NImmMaxNarrow{} (narrow) and \NImmMaxNB{} (near-bamboo) of the lines are immortal in any condition. Wide lines have only the lognormal multiplier contrasts. Without Cu they live longer than with blocking ends and many become immortal at low current density: the immortal fraction is \NImmWideResCuZeroJOne{} at 1~\MAcm{} and \NImmWideResCuZeroJTwo{} at 2~\MAcm. With 0.5 or 2~wt\% Cu, no wide line is immortal at any current density, and at 3~\MAcm{} the median is \NTWideResCuHalfJThree~h (0.5~wt\%) against \NTWideBlkCuHalfJThree~h with blocking ends. Copper redistribution creates the flux divergences that the wide microstructure lacks. In the seed-0 fields every wide reservoir-end void forms within 2~\um{} of a step in the Cu content, from \NSiteCuMinWideResMin--\NSiteCuMinWideResMax~$c_0$ on one side to \NSiteCuMaxWideResMin--\NSiteCuMaxWideResMax~$c_0$ on the other, and through $\kappa_{\mathrm{eff}} = \kappa/(1+\beta c)$ such a step is a $\kappa$ contrast of an order of magnitude. By the time of nucleation the capped Cu field has steps along much of the line, so the sign of the step is the better test. Rebuilding the Cu field at $t_{\mathrm{nuc}}$ for \NStepRuns{} of these runs (the first 100 maps at each Cu content and current density), the Cu falls in the direction of Al drift across the void site in \NStepFallPct\% of them, with a median ratio of downstream to upstream Cu of \NStepSignedMed, against \NStepFallRandPct\% at random positions on the same lines; the Cu contrast at the site exceeds that at \NStepRankPct\% of the positions on its line (median). A Cu-rich, slow segment upstream of a Cu-poor, fast one removes Al faster than it is supplied, which is where tension builds. Section~\ref{sec:res:cutransport} shows that this needs both Cu transport and the immobile precipitate. Only \NHillockWideRes\% of the wide reservoir-end runs reach the hillock threshold, and no blocking-end run does. The early-failure tail of this class is $t_{01} = \NTzeroOneRatioWideRes$ (\NTzeroOneRatioWideResLo--\NTzeroOneRatioWideResHi) of the median at 0.5~wt\% and 3~\MAcm, as for the other classes.

The near-bamboo class is bimodal at 1~\MAcm. Some lines fail early while others fail only after long transients, and $\sigma_{\ln}$ reaches \NNBJOneSigMin{} to \NNBJOneSigMax. The lognormal medians (\NNBJOneLognMin{} to \NNBJOneLognMax~h) then differ from the Kaplan-Meier medians (\NNBJOneKMMin{} to \NNBJOneKMMax~h) by up to an order of magnitude. Above 1~\MAcm{} the near-bamboo distributions are narrow again ($\sigma_{\ln} = \NSigBlkNBJgeMin$ to \NSigBlkNBJgeMax{} at 3 to 6~\MAcm{} with blocking ends). Knowlton found the same kind of bimodality in near-bamboo populations.\cite{knowltonthesis}

\emph{Sensitivity to compressive relief.} No blocking-end run reaches $\sigma_{\mathrm{cc}}$, and the blocking-end nucleation times without relief are identical to those with it (largest relative difference \NNoReliefBlkMaxRel). With reservoir ends, on the same \NNoReliefMaps{} wide-line maps at 2~\MAcm{} and 0.5~wt\% Cu, the immortal fraction is \NNoReliefImmWideResJTwo{} without relief and \NReliefImmWideResJTwoSame{} with it, and the minimum stress without relief is \NSigMinNoRelief~GPa. With the solubility cap the results therefore do not depend on the relief model.

\begin{table}
\caption{Median void-nucleation time $t_{50}$ (h) and lognormal spread $\sigma_{\ln}$ at $j = 3$~\MAcm, 200~$^{\circ}$C and $L = 280$~\um, with \NEthreeMaps{} maps per condition; $f_{\mathrm{imm}}$ is the immortal fraction with reservoir ends (it is zero with blocking ends).}
\label{tab:t50}
\begin{ruledtabular}
\begin{tabular}{llccccc}
 & & \multicolumn{2}{c}{Blocking} & \multicolumn{3}{c}{Reservoir} \\
Class & Cu (wt\%) & $t_{50}$ & $\sigma_{\ln}$ & $t_{50}$ & $\sigma_{\ln}$ & $f_{\mathrm{imm}}$ \\
\hline
Wide & 0 & \NTWideBlkCuZeroJThree & \NSigWideBlkCuZeroJThree & \NTWideResCuZeroJThree & \NSigWideResCuZeroJThree & \NImmWideResCuZeroJThree \\
 & 0.5 & \NTWideBlkCuHalfJThree & \NSigWideBlkCuHalfJThree & \NTWideResCuHalfJThree & \NSigWideResCuHalfJThree & \NImmWideResCuHalfJThree \\
 & 2 & \NTWideBlkCuTwoJThree & \NSigWideBlkCuTwoJThree & \NTWideResCuTwoJThree & \NSigWideResCuTwoJThree & \NImmWideResCuTwoJThree \\
Narrow & 0 & \NTNarrowBlkCuZeroJThree & \NSigNarrowBlkCuZeroJThree & \NTNarrowResCuZeroJThree & \NSigNarrowResCuZeroJThree & \NImmNarrowResCuZeroJThree \\
 & 0.5 & \NTNarrowBlkCuHalfJThree & \NSigNarrowBlkCuHalfJThree & \NTNarrowResCuHalfJThree & \NSigNarrowResCuHalfJThree & \NImmNarrowResCuHalfJThree \\
 & 2 & \NTNarrowBlkCuTwoJThree & \NSigNarrowBlkCuTwoJThree & \NTNarrowResCuTwoJThree & \NSigNarrowResCuTwoJThree & \NImmNarrowResCuTwoJThree \\
Near-bamboo & 0 & \NTNBBlkCuZeroJThree & \NSigNBBlkCuZeroJThree & \NTNBResCuZeroJThree & \NSigNBResCuZeroJThree & \NImmNBResCuZeroJThree \\
 & 0.5 & \NTNBBlkCuHalfJThree & \NSigNBBlkCuHalfJThree & \NTNBResCuHalfJThree & \NSigNBResCuHalfJThree & \NImmNBResCuHalfJThree \\
 & 2 & \NTNBBlkCuTwoJThree & \NSigNBBlkCuTwoJThree & \NTNBResCuTwoJThree & \NSigNBResCuTwoJThree & \NImmNBResCuTwoJThree \\
\end{tabular}
\end{ruledtabular}
\end{table}

\subsection{Nucleation sites (E3)}

\begin{figure}[t]
\includegraphics[width=\columnwidth]{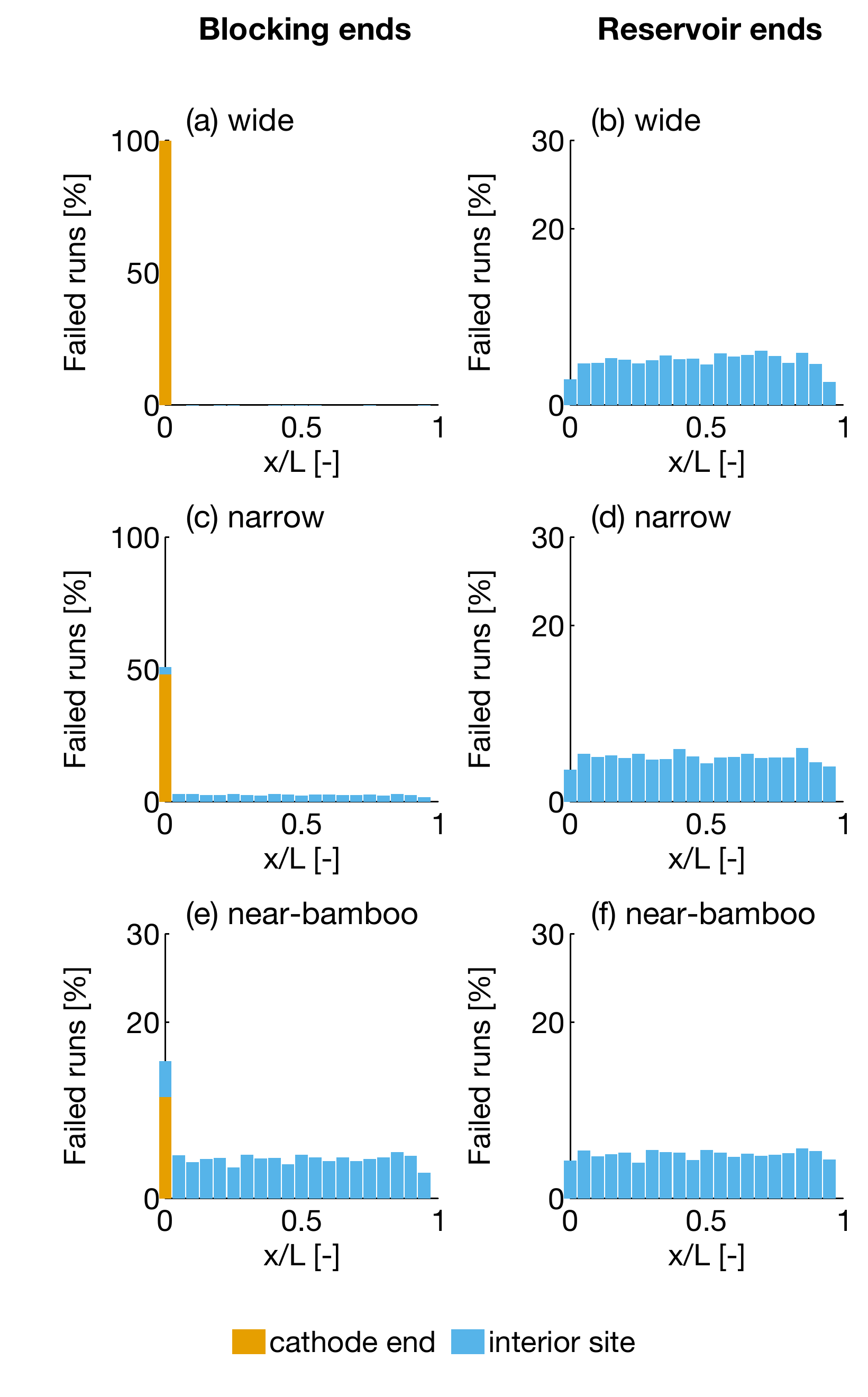}
\caption{Distribution of the void-nucleation site $x_{\mathrm{nuc}}/L$ for failed runs, pooled over Cu content (0, 0.5 and 2~wt\%) and $j = 1$ to 6~\MAcm{} (\NEthreeMaps{} maps per condition), for wide, narrow and near-bamboo lines with blocking (left) and reservoir (right) ends. Bars give the percentage of failed runs in each bin of width $0.05L$, split into cathode-end voids (orange) and interior voids (blue). With blocking ends the cathode share is \NCathWideBlk\% (wide), \NCathNarrowBlk\% (narrow) and \NCathNBBlk\% (near-bamboo); with reservoir ends every void is interior. Note the different vertical scales.}
\label{fig:location}
\end{figure}

Figure~\ref{fig:location} shows where the failed lines nucleate, pooled over Cu content and current density. With blocking ends, every wide line fails at the cathode (\NCathWideBlk\%). The cathode layer reaches $\sigma_{\mathrm{crit}}$ before any interior site, because the interior divergences of wide lines come only from the multiplier contrasts and from Cu redistribution, which takes longer. In narrow lines \NCathNarrowBlk\% of the voids form at the cathode and the rest in the interior; \NLtTenNarrowBlk\% lie within $x/L < 0.1$, \NLtHalfNarrowBlk\% in the cathode half, and \NGtNineNarrowBlk\% within $x/L > 0.9$. In near-bamboo lines only \NCathNBBlk\% form at the cathode and \NLtHalfNBBlk\% in the cathode half. Of all interior voids, \NTdivShare\% sit at tensile flux divergences of the microstructure, boundaries where $\kappa$ increases in the direction of the electron flow. With reservoir ends the end stress is pinned, every void is interior, and the sites are spread along the line, with \NResHalfMin\% to \NResHalfMax\% in the cathode half depending on the class.

\subsection{Apparent Black's-law exponent (E5)}
\label{sec:res:e5}

\begin{figure}[t]
\includegraphics[width=\columnwidth]{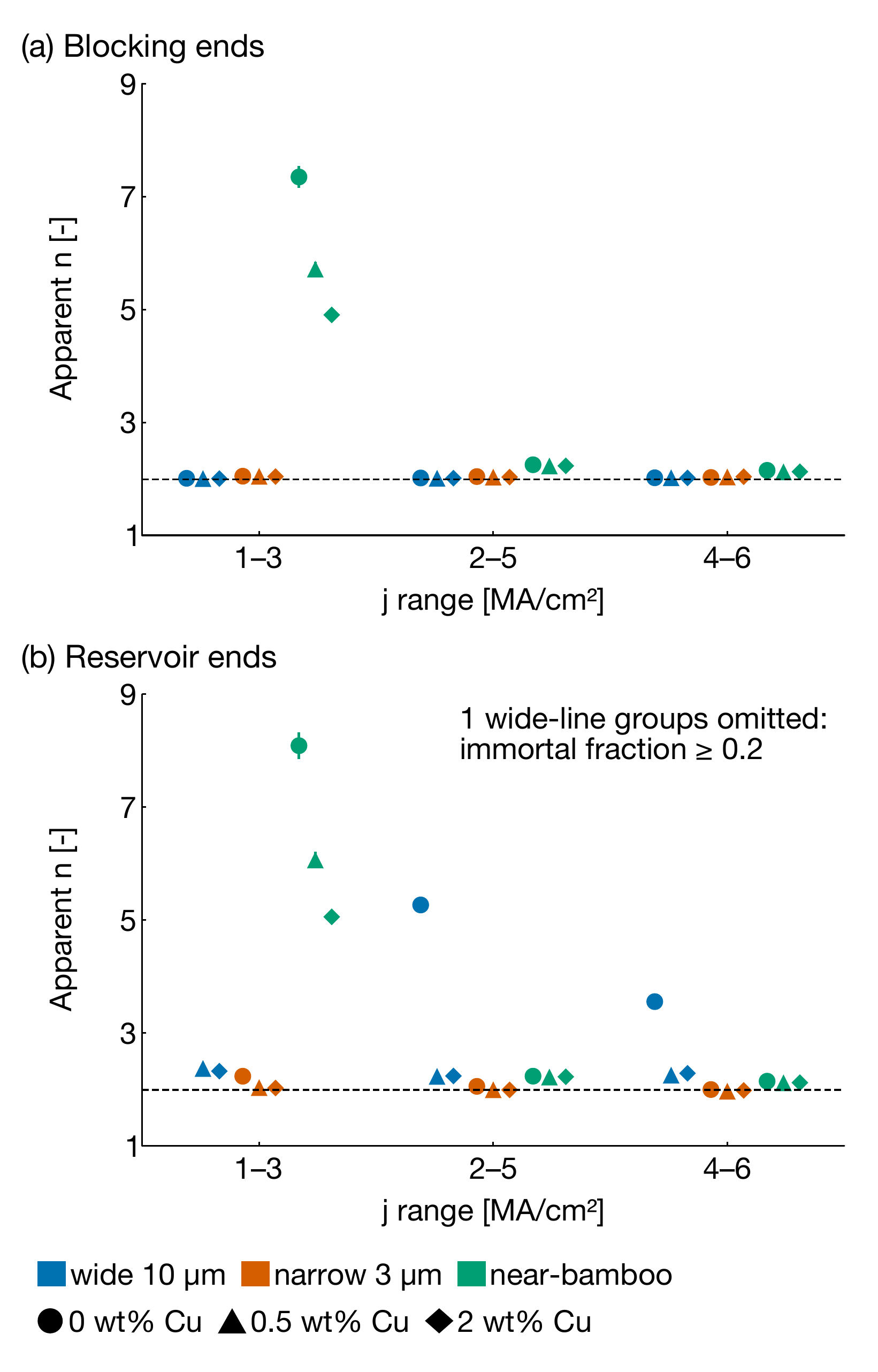}
\caption{Apparent current-density exponent $n$ of Black's law, fitted by a censored lognormal accelerated-failure-time model to the per-run lifetimes over the current-density ranges 1 to 3, 2 to 5 and 4 to 6~\MAcm, for (a) blocking and (b) reservoir ends. Color gives the width class and the symbol the Cu content; bars are 95\% confidence intervals and are hidden by the symbol where shorter than it. The dashed line marks $n = 2$. For \NBnGated{} reservoir-end group, Cu-free wide lines at 1 to 3~\MAcm, $n$ is not reported (immortal fraction of \NImmGate{} or more in at least one condition of the range).}
\label{fig:blackn}
\end{figure}

The fitted exponents are in Fig.~\ref{fig:blackn}. With blocking ends we get $n = \NBnWideBlkMin$ to \NBnWideBlkMax{} for wide lines and \NBnNarrowBlkMin{} to \NBnNarrowBlkMax{} for narrow ones; no 95\% interval is wider than $\pm\NBnCIHalfMax$. Adding Cu moves the exponent by \NBnCuShiftMax{} at most, for the same class and current-density range. So when every line fails and most failures sit at the cathode, Cu leaves $n$ alone. What it changes is the prefactor, and the activation energy through the solvus and the Cu activation energy (Sec.~\ref{sec:res:e1}). This agrees with Park \emph{et al.}\ and with Lloyd and Clement, who found $n = 2$ for nucleation-limited failure in Al-Cu.\cite{park1999,lloyd1996}

The near-bamboo class, where failure is controlled by interior divergences at bamboo grains, gives $n = \NBnNBJgeMin$ to \NBnNBJgeMax{} above 1~\MAcm, where nearly every line fails, and Cu does not change it: with blocking ends over 2 to 5~\MAcm{} it is \NBnNBBlkCuZeroJTwoFive, \NBnNBBlkCuHalfJTwoFive{} and \NBnNBBlkCuTwoJTwoFive{} for 0, 0.5 and 2~wt\% Cu. Narrow lines with reservoir ends, which also fail only at interior divergences, give \NBnNarrowResCuZeroJTwoFive, \NBnNarrowResCuHalfJTwoFive{} and \NBnNarrowResCuTwoJTwoFive. The Cu-free near-bamboo exponent is itself slightly above 2. With all Cu mobile, the near-bamboo exponent rose with Cu content, to \NTenNMobNBBlkCuTwoJTwoFive{} at 2~wt\%; this rise disappears once only the dissolved Cu moves (Sec.~\ref{sec:res:cutransport}).

Larger exponents arise in two situations. The first is a population that is no longer a single nucleation-limited mode. Near-bamboo lines over 1 to 3~\MAcm{} give $n = \NBnNBJOneMin$ to \NBnNBJOneMax, from the bimodal distribution at 1~\MAcm; here Cu lowers $n$, from \NBnNBBlkCuZeroJOneThree{} to \NBnNBBlkCuTwoJOneThree{} with blocking ends. The second is the Cu-free wide line with reservoir ends, whose stress builds only at the weak multiplier contrasts. Over 4 to 6~\MAcm, where every such line fails, it gives $n = \NBnWideResCuZeroJFourSix$; over 2 to 5~\MAcm{} it gives \NBnWideResCuZeroJTwo, with a largest immortal fraction of \NBnWideResCuZeroJTwoImm{} in the range; and over 1 to 3~\MAcm{} the immortal fraction reaches \NImmGate{} and $n$ is not reported. These departures from $n = 2$ come from bimodality and from immortality or the approach to it, the causes that Knowlton identified without solute.\cite{knowltonthesis} Copper removes the second one: with 0.5 or 2~wt\% Cu the wide reservoir-end lines give $n = \NBnWideResCuHalfJTwoFive$ to \NBnWideResCuHalfJOneThree{} over every range, close to 2, because they now fail at Cu steps (Sec.~\ref{sec:res:e3}).

\subsection{Immortality and the Blech product (E4)}

\begin{figure}[t]
\includegraphics[width=\columnwidth]{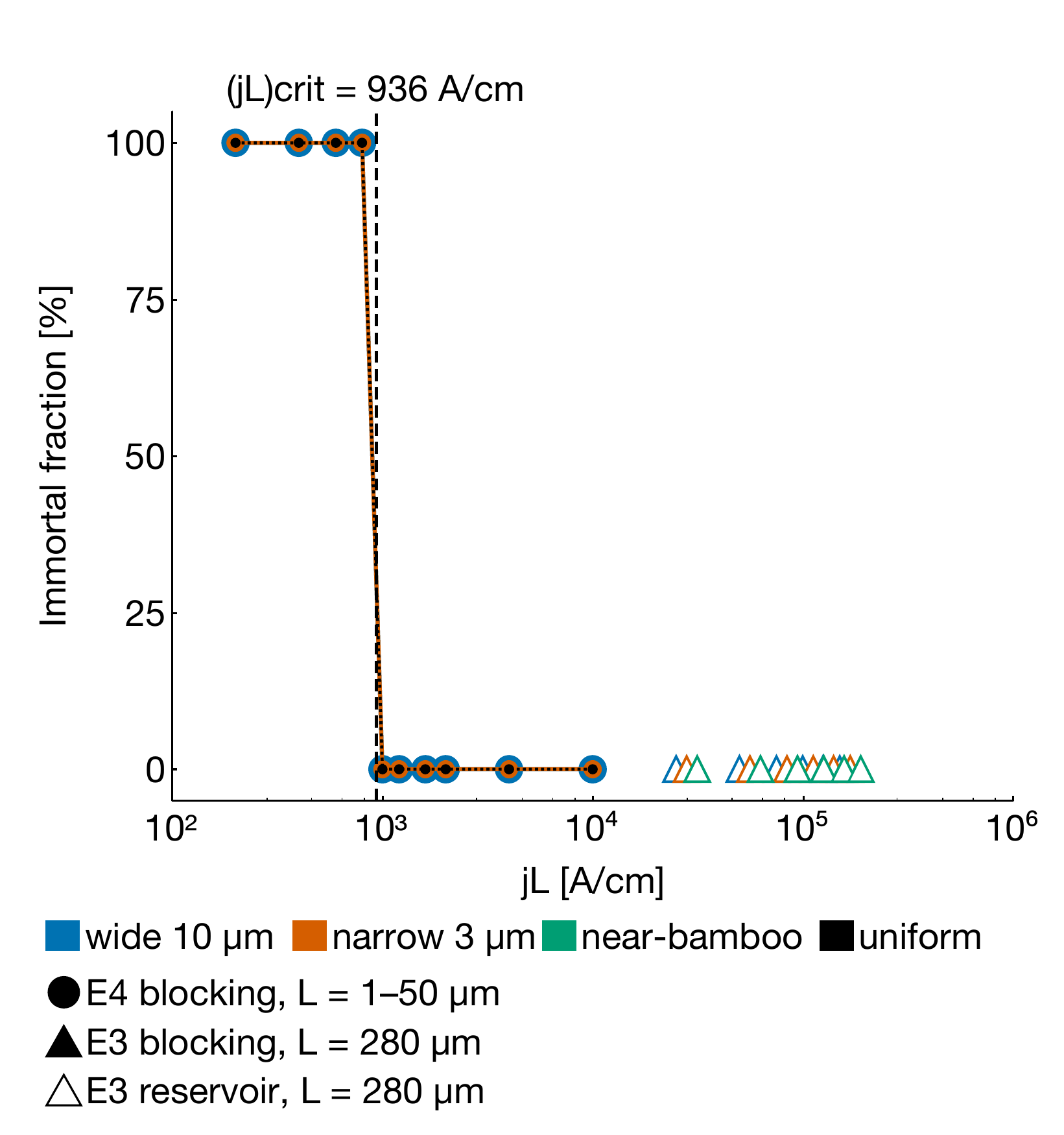}
\caption{Immortal fraction (\%) against the product $jL$ (A/cm). Lines: Al-0.5~wt\% Cu lines with blocking ends at $j = 2$~\MAcm{} and $L = 1$ to 50~\um{} (\NEfourMaps{} maps per class for wide and narrow lines, one run for the uniform line; all Cu mobile). Symbols: the $L = 280$~\um{} lines at $j = 1$ to 6~\MAcm{} (\NEthreeMaps{} maps per condition) with blocking (filled) and reservoir (open) ends; these symbols are offset horizontally by class (wide left, near-bamboo right, 0.05 decade) for visibility. The dashed vertical line is the analytic Blech product $(jL)_{\mathrm{crit}} = \NJLcrit$~A/cm ($L_{\mathrm{crit}} = \NLcrit$~\um{} at 2~\MAcm). Every blocking-end line above $(jL)_{\mathrm{crit}}$ fails; immortality far above $(jL)_{\mathrm{crit}}$ occurs only in Cu-free wide lines with reservoir ends (immortal fractions of \NImmWideResCuZeroJOne{} and \NImmWideResCuZeroJTwo{} at $j = 1$ and 2~\MAcm).}
\label{fig:blech}
\end{figure}

With blocking ends the zero-flux steady state has $\partial_x\sigma = -G$ wherever $\kappa_{\mathrm{eff}} > 0$, whatever the microstructure and the Cu profile, so the cathode stress tends to $\sigma_{\mathrm{res}} + GL/2$ in every line. The Blech threshold of a blocking-end line is therefore $(jL)_{\mathrm{crit}} = 2(\sigma_{\mathrm{crit}} - \sigma_{\mathrm{res}})\Omega/(e\Zs_{\mathrm{Al}}\rho) = \NJLcrit$~A/cm, or $L_{\mathrm{crit}} = \NLcrit$~\um{} at 2~\MAcm, independent of the class and of how the Cu is distributed. E4 confirms this (Fig.~\ref{fig:blech}). Every wide, narrow and uniform line with $L \le \NEfourLImm$~\um{} is immortal and every line with $L \ge \NEfourLFail$~\um{} fails. Because the threshold does not depend on the Cu profile, E4 is run with all Cu mobile (Sec.~\ref{sec:methods}); a check with the solubility cap on \NEfourCapMaps{} maps per class at $L = 3$ to 6~\um{} gives the same threshold, with every line at $L \le \NEfourCapLImm$~\um{} immortal and every line at $L \ge \NEfourCapLFail$~\um{} failing. Immortality far above $(jL)_{\mathrm{crit}}$ occurs only in Cu-free wide lines with reservoir ends: at $L = 280$~\um, \NImmWideResCuZeroJOne{} of them are immortal at 1~\MAcm{} and \NImmWideResCuZeroJTwo{} at 2~\MAcm, for $jL = 2.8\times10^{4}$ and $5.6\times10^{4}$~A/cm. This is the same kind of pad-terminated immortality at low $j$ that Knowlton reported for near-bamboo lines.\cite{knowltonthesis} With Cu, every such line fails.

\subsection{How much the Cu moves (E7, E10)}
\label{sec:res:cutransport}

The main model lets only the dissolved Cu move. Two limiting cases are run on the same maps. In E7 the Cu is held immobile: the solute flux is zero on every face, so $c$ stays at $c_0$, while $\kappa_{\mathrm{eff}} = \kappa/(1 + \beta c_0)$ still applies on grain-boundary cells; the scheduled horizon is lengthened to $(1 + \beta c_0)\,t_{\mathrm{end}}$ so that every immobile run that does not nucleate is certified steady (\NEsevenTendStops{} runs stop at the horizon). In E3\textsubscript{mobile} all Cu is treated as mobile solute (no cap), the model of earlier solute-coupled work. Both controls use the E3 conditions with 0.5 and 2~wt\% Cu on the first \NEsevenMaps{} maps of each class (\NEsevenRuns{} runs each). E10 adds, on the same maps, the capped model with blocking by the dissolved Cu only (\NTenRuns{} runs, with the main model repeated as a check), and E11 lets a fraction $f$ of the precipitate block (\NElevenRuns{} runs). Ratios of medians are given with paired bootstrap 95\% intervals.

\emph{Cathode-dominated lines.} Immobile Cu acts as a static blocking factor. For wide lines with blocking ends at 3~\MAcm{} it raises the median over the Cu-free value by \NSevenStaticCuHalf{} and \NSevenStaticCuTwo{} at 0.5 and 2~wt\%, equal to $1 + \beta c_0$ (\PBlockHalf{} and \PBlockTwo). With the solubility cap the median is slightly lower: in wide and narrow lines with blocking ends the immobile median is \NSevenRatioBlkWNCuHalfMin{} to \NSevenRatioBlkWNCuHalfMax{} times the capped one at 0.5~wt\% and \NSevenRatioBlkWNCuTwoMin{} to \NSevenRatioBlkWNCuTwoMax{} times at 2~wt\%, at every current density. With all Cu mobile it is much lower: the capped median is \NTenRatioTotBlkWNCuHalfMin{} to \NTenRatioTotBlkWNCuHalfMax{} times the all-mobile one at 0.5~wt\% and \NTenRatioTotBlkWNCuTwoMin{} to \NTenRatioTotBlkWNCuTwoMax{} times at 2~wt\%. For wide lines at 3~\MAcm{} and 2~wt\% the three medians are \NSevenTImmWideBlkCuTwoJThree~h (immobile), \NTenTTotWideBlkCuTwoJThree~h (capped) and \NTenTMobWideBlkCuTwoJThree~h (all mobile). Treating all Cu as mobile thus overstates how much Cu leaves the cathode layer, by an amount that grows with the Cu content, because the real drift is limited to the dissolved fraction, \PMobFracHalf{} of the Cu at 0.5~wt\% and \PMobFracTwo{} at 2~wt\%. None of these ratios depends on $j$, so in these lines the three models give the same exponent to within \NTenDnBlkWNMax.

\emph{Wide lines with reservoir ends.} Here the three models differ qualitatively. With immobile Cu the immortal fraction is \NSevenImmImmCuHalfJOne{} at 1~\MAcm{} and \NSevenImmImmCuHalfJTwo{} at 2~\MAcm, the same as without Cu (\NImmWideResCuZeroJOne{} and \NImmWideResCuZeroJTwo), because a uniform, immobile Cu content scales $\kappa$ uniformly, as for Mn in Sec.~\ref{sec:res:cumn}. With all Cu mobile it rises to \NTenImmMobCuHalfJOne{} and \NTenImmMobCuHalfJTwo{} at 0.5~wt\%: Cu piles up toward the anode and makes the cathode part of the line compressive. That pile-up reaches tens of at.\%, far above the solubility, so the immortality it produces is an artifact of treating all Cu as mobile. With the cap no line is immortal (\NTenImmTotCuHalfJOne{} and \NTenImmTotCuHalfJTwo), and where every line fails, at 4 to 6~\MAcm, the immobile median is \NSevenRatioWideResMin{} to \NSevenRatioWideResMax{} times the capped one. The capped model fails these lines at the Cu steps of Sec.~\ref{sec:res:e3}; with all Cu mobile, back-diffusion of the dissolved Cu smooths such steps, and with immobile Cu they never form. The exponent follows: over 4 to 6~\MAcm{} it is \NBnWideResCuZeroJFourSix{} without Cu and \NSevenNImmWideResCuHalfJFourSix{} with immobile Cu, \NTenNMobWideResCuHalfJFourSix{} and \NTenNMobWideResCuTwoJFourSix{} with all Cu mobile, and \NTenNTotWideResCuHalfJFourSix{} and \NTenNTotWideResCuTwoJFourSix{} with the cap, at 0.5 and 2~wt\%.

\emph{Near-bamboo lines.} Above 1~\MAcm{} the immobile median is \NSevenRatioNBCuHalfMin{} to \NSevenRatioNBCuHalfMax{} times the capped one at 0.5~wt\% and \NSevenRatioNBCuTwoMin{} to \NSevenRatioNBCuTwoMax{} times at 2~wt\%, and the capped median is \NTenRatioTotNBCuHalfMin{} to \NTenRatioTotNBCuHalfMax{} and \NTenRatioTotNBCuTwoMin{} to \NTenRatioTotNBCuTwoMax{} times the all-mobile one. At 1~\MAcm, where the distributions are bimodal, the immobile median is \NSevenRatioNBJOneMin{} to \NSevenRatioNBJOneMax{} times the capped one. With blocking ends over 2 to 5~\MAcm{} the exponent is \NSevenNImmNBBlkCuHalfJTwoFive{} and \NSevenNImmNBBlkCuTwoJTwoFive{} with immobile Cu, \NTenNTotNBBlkCuHalfJTwoFive{} and \NTenNTotNBBlkCuTwoJTwoFive{} with the cap, and \NTenNMobNBBlkCuHalfJTwoFive{} and \NTenNMobNBBlkCuTwoJTwoFive{} with all Cu mobile, against \NBnNBBlkCuZeroJTwoFive{} without Cu. The rise of $n$ with Cu content that the all-mobile model gives is thus another consequence of moving more Cu than is dissolved.

\emph{Blocking by dissolved Cu only.} If precipitates did not block grain-boundary diffusion, only the dissolved Cu would enter $\kappa_{\mathrm{eff}}$. The lifetime gain from Cu would then be capped by $1 + \beta c_{\mathrm{sol}}$ and would not grow with the Cu content: for wide lines with blocking ends at 3~\MAcm{} the median is \NTenTDisWideBlkCuHalfJThree~h at 0.5~wt\% and \NTenTDisWideBlkCuTwoJThree~h at 2~wt\%, \NTenRatioDisBlkWNCuHalfMin{} to \NTenRatioDisBlkWNCuHalfMax{} and \NTenRatioDisBlkWNCuTwoMin{} to \NTenRatioDisBlkWNCuTwoMax{} times the all-mobile median. With reservoir ends this variant gives wide lines an immortal fraction of \NTenImmDisCuHalfJOne{} (0.5~wt\%) and \NTenImmDisCuTwoJOne{} (2~wt\%) at 1~\MAcm. Our blocking constant $\beta$ was calibrated on the total Cu content (Sec.~\ref{sec:model}), which is why the main model uses the total Cu.

\emph{Partial blocking by precipitate.} How strongly precipitates block grain-boundary diffusion is not known, so E11 interpolates between the two variants: the Cu that blocks is $c_m + f\,(c - c_m)$, the dissolved Cu plus a fraction $f$ of the precipitate, with $f = 0.25$, 0.5 and 0.75 on the E10 maps ($f = 1$ is the main model and $f = 0$ the dissolved-only variant). In wide and narrow lines with blocking ends the median falls smoothly as $f$ decreases, to \NElevenRatioThreeQCuHalfMin--\NElevenRatioThreeQCuHalfMax, \NElevenRatioHalfCuHalfMin--\NElevenRatioHalfCuHalfMax{} and \NElevenRatioQuarterCuHalfMin--\NElevenRatioQuarterCuHalfMax{} of the main-model value at $f = 0.75$, 0.5 and 0.25 with 0.5~wt\% Cu, and to \NElevenRatioThreeQCuTwoMin--\NElevenRatioThreeQCuTwoMax, \NElevenRatioHalfCuTwoMin--\NElevenRatioHalfCuTwoMax{} and \NElevenRatioQuarterCuTwoMin--\NElevenRatioQuarterCuTwoMax{} with 2~wt\%. The Cu-step failures of wide lines with reservoir ends, by contrast, do not need full blocking: at every $f \ge 0.25$ the immortal fraction at 1 and 2~\MAcm{} is \NElevenImmPartMax, and only at $f = 0$ does the immortality return (\NElevenImmZeroJOneMin{} to \NElevenImmZeroJOneMax{} at 1~\MAcm). Over 4 to 6~\MAcm{} the exponent of these lines is \NElevenNWideResQuarterMin{} to \NElevenNWideResQuarterMax{} at $f = 0.25$, against \NElevenNWideResZeroMin{} to \NElevenNWideResZeroMax{} at $f = 0$. The near-bamboo exponent over 2 to 5~\MAcm{} is \NElevenNNBMin{} at every $f$. With $f = 0$ the median nucleation time at 2~wt\% Cu is \NElevenCuRatioZeroMin{} to \NElevenCuRatioZeroMax{} times that at 0.5~wt\% (geometric mean over $j$), so Cu above the solubility adds almost nothing; with $f \ge 0.25$ the factor is \NElevenCuRatioPartMin{} to \NElevenCuRatioPartMax. So $f$ sets how much Cu helps, but every $f$ we ran above zero removes the immortality of these wide lines, as the main model does.

\subsection{Temperature and extrapolation (E9)}
\label{sec:res:e9}

E9 repeats the E3 conditions at 150, 250 and 300~$^{\circ}$C on the first \NEnineMaps{} maps of each class (\NEnineRuns{} runs); at 200~$^{\circ}$C the same maps are taken from E3. Over 150 to 300~$^{\circ}$C and 2 to 5~\MAcm, the apparent activation energy of the populations with blocking ends is \NNineEaBlkMin{} to \NNineEaBlkMax~eV and rises with Cu content, as in the uniform line (Sec.~\ref{sec:res:e1}). With Cu the lifetime is not Arrhenius: for wide lines with blocking ends at 2~\MAcm{} and 2~wt\% Cu the local activation energy is \NNineEaLocCuTwoLo~eV between 150 and 200~$^{\circ}$C and \NNineEaLocCuTwoHi~eV between 250 and 300~$^{\circ}$C, against \NNineEaLocCuZeroLo{} and \NNineEaLocCuZeroHi~eV without Cu. The dissolved fraction grows with temperature along the solvus, so at high temperature a larger part of the Cu can leave the cathode layer, and the lifetime falls faster with $T$ there than at low temperature. In near-bamboo lines the exponent fitted over 1 to 3~\MAcm{} at all four temperatures is \NNineNBnLowMin{} to \NNineNBnLowMax: the long-lived part of the population at 1~\MAcm{} (Sec.~\ref{sec:res:e3}) is present at every temperature.

\begin{figure}[t]
\includegraphics[width=\columnwidth]{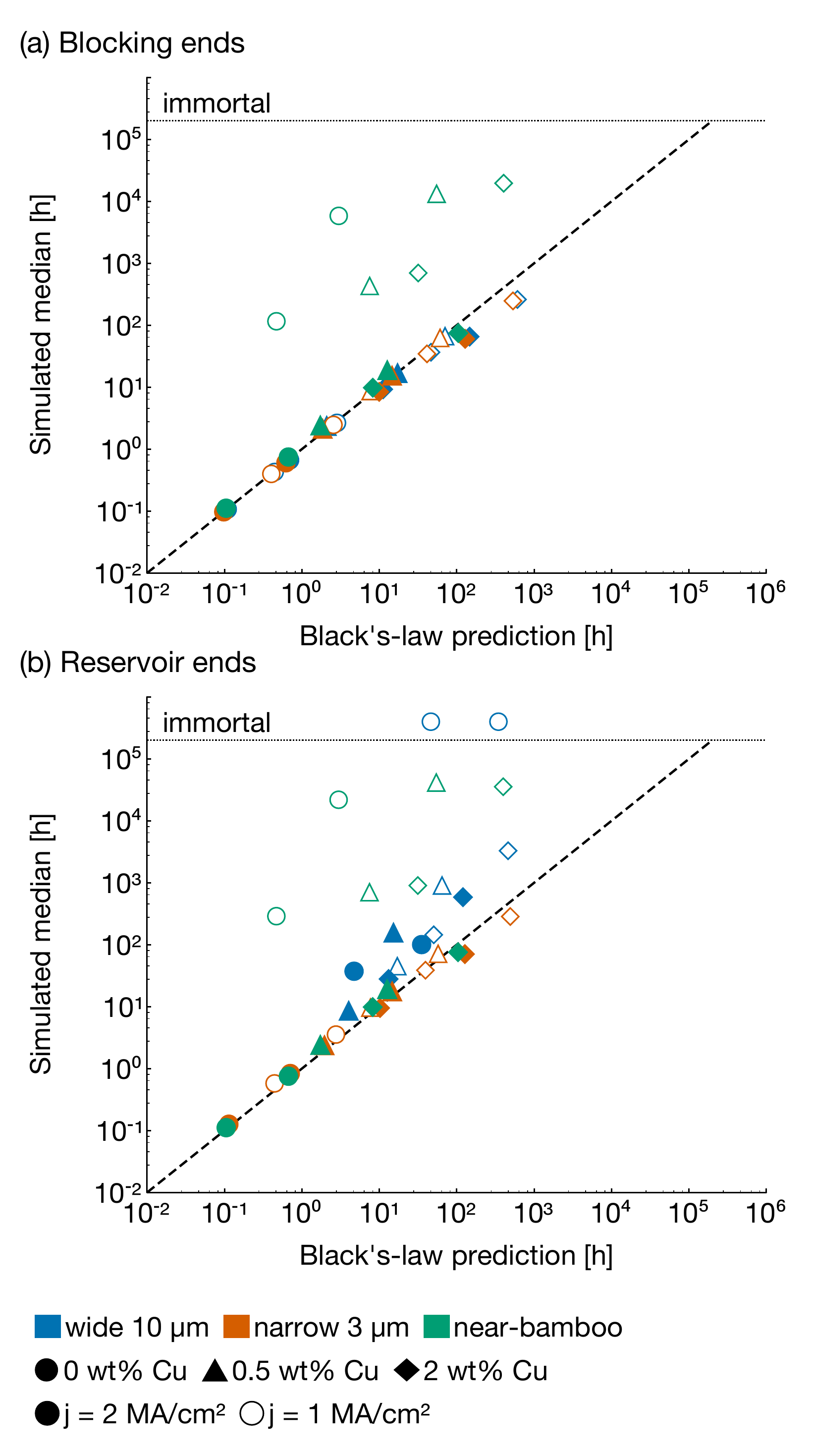}
\caption{Extrapolation from accelerated tests (E9). Median nucleation time predicted by Black's law against the simulated median at 150 and 200~$^{\circ}$C and 1 and 2~\MAcm, for (a) blocking and (b) reservoir ends. For each class, Cu content and end type, $n$, $E_a$ and the prefactor are fitted by a censored lognormal accelerated-failure-time model to the runs at \NNineAccTlo{} and \NNineAccThi~$^{\circ}$C and 4 to 6~\MAcm. Color gives the width class and the symbol the Cu content; filled symbols are $j = 2$ and open symbols $j = 1$~\MAcm. Populations whose Kaplan-Meier median is infinite are drawn in the band marked ``immortal'' at their predicted median. The dashed line marks prediction equal to simulation; points above it are lifetimes the extrapolation underestimates, and points below it lifetimes it overestimates (all at 2~wt\% Cu).}
\label{fig:extrap}
\end{figure}

E9 also tests the step by which lifetimes from accelerated tests are carried to use conditions with the exponent and activation energy fitted at the test conditions.\cite{sapatnekar2026,tanimoto2010} For each class, Cu content and end type we fit $n$, $E_a$ and the prefactor (censored lognormal AFT, Sec.~\ref{sec:stats}) to the runs at \NNineAccTlo{} and \NNineAccThi~$^{\circ}$C and 4 to 6~\MAcm, predict the median at 150 and 200~$^{\circ}$C and 1 and 2~\MAcm{} with Black's law, and compare the prediction with the median simulated at that condition (Fig.~\ref{fig:extrap}). In wide and narrow lines with blocking ends, which fail at the cathode with $n = \NNinenAccBlkWNMin$ to \NNinenAccBlkWNMax{} at the test conditions, the prediction is within a factor of \NNineRatioBlkWNLowCuMax{} of the simulated median at 0 and 0.5~wt\% Cu. At 2~wt\% it is too long: \NNineRatioBlkWNCuTwoOneFiftyMin{} to \NNineRatioBlkWNCuTwoOneFiftyMax{} times the simulated median at 150~$^{\circ}$C and \NNineRatioBlkWNCuTwoTwoHundredMin{} to \NNineRatioBlkWNCuTwoTwoHundredMax{} times at 200~$^{\circ}$C, because the activation energy fitted at the test temperatures is the high-temperature one. In narrow lines with reservoir ends the prediction is \NNineRatioNarrowResMin{} to \NNineRatioNarrowResMax{} times the simulated median, the upper value again at 2~wt\% and 150~$^{\circ}$C. Near-bamboo lines ($n = \NNinenAccNBMin$ to \NNinenAccNBMax{} at the test conditions) are predicted at \NNineRatioNBJtwoMin{} to \NNineRatioNBJtwoMax{} times the simulated median at 2~\MAcm, but at 1~\MAcm{} the simulated median is \NNineNBJoneFacMin{} to \NNineNBJoneFacMax{} times longer than the prediction. Wide lines with reservoir ends give $n = \NNinenAccWideResMin$ to \NNinenAccWideResMax{} at the test conditions; at the use conditions \NNineWideResImmUse{} of the \NNineWideResUse{} populations (all Cu-free) have an infinite Kaplan-Meier median, and where the median is finite the prediction is \NNineRatioWideResMin{} to \NNineRatioWideResMax{} times it. Of the \NNineCount{} predictions, \NNineOverCount{} exceed the simulated median by more than \NNineOverTol\%, all of them at 2~wt\% Cu; the largest is a factor of \NNineRatioMax. Where Cu is present well above its solubility, Black's-law extrapolation from high-temperature tests can therefore overestimate the lifetime.

\subsection{Application to Cu(Mn) lines (E6)}
\label{sec:res:cumn}

\begin{figure}[t]
\includegraphics[width=\columnwidth]{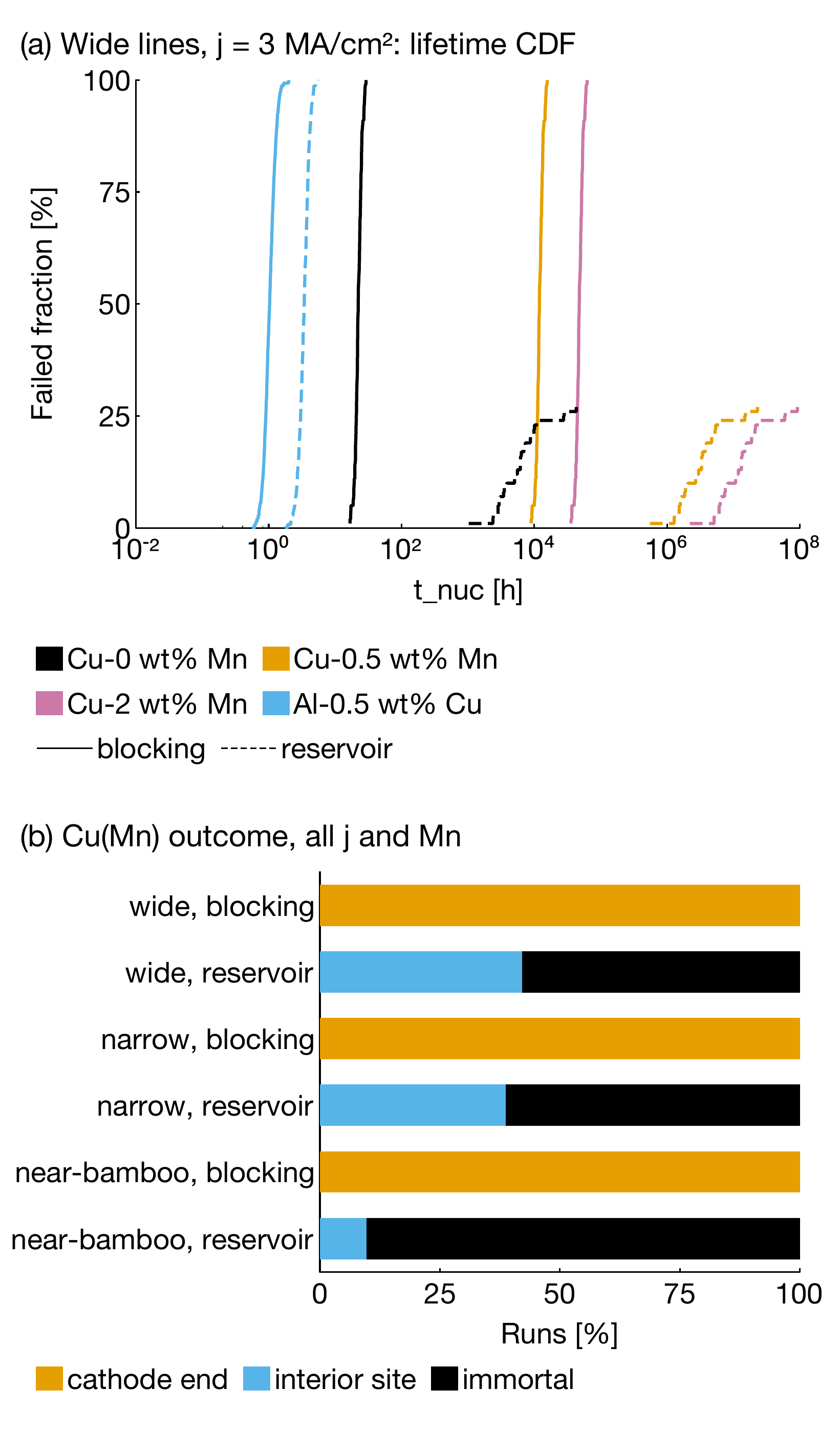}
\caption{Application to Cu(Mn) lines. (a) Fraction of lines failed vs. nucleation time $t_{\mathrm{nuc}}$ (Kaplan-Meier estimate) at $j = 3$~\MAcm, for wide Cu lines with 0, 0.5 and 2~wt\% Mn (\NEsixMaps{} maps each) and, for reference, wide Al-0.5~wt\% Cu lines (\NEthreeMaps{} maps). Solid curves: blocking ends; dashed curves: reservoir ends. Lines that never nucleate a void (immortal lines) count as surviving, so the reservoir-end Cu(Mn) curves level off near \NCuMnFailWide\%. (b) Share of Cu(Mn) runs ending in a void at the cathode end, a void in the interior, or no void (immortal), vs. width class and end type, pooled over Mn content and $j = 1$ to 6~\MAcm{} (1800 runs per bar).}
\label{fig:cumn}
\end{figure}

To check that the solver is not specific to Al-Cu, we ran Cu lines with 0, 0.5 and 2~wt\% Mn using the Cu parameter set of Alam \emph{et al.}\cite{alam2005,alshohel2021} (Table~\ref{tab:params}), \NEsixMaps{} maps per condition and the E3 grid of classes, current densities and ends (\NEsixRuns{} runs). Every grain-boundary and bamboo cell carries interface-dominated Cu diffusion. Manganese is treated as immobile: its lattice diffusivity\cite{butrymowicz1975} is negligible on the test time scale, and no grain-boundary or interface data were found. Its blocking constant $\beta_{\mathrm{Mn}} = \PBetaMn$ is derived from the \PDeltaEaMn~eV rise of the EM activation energy reported by Hu \emph{et al.};\cite{hu2012} depending on the temperature at which that rise is evaluated it lies between \PBetaMnLow{} and \PBetaMnHigh, and we use the upper value. The resistivity is held at its quoted value. With blocking ends every Cu(Mn) line fails at the cathode. For wide lines at 3~\MAcm{} the median is \NCuMnTZero~h without Mn, \NCuMnTHalf~h with 0.5~wt\% and \NCuMnTTwo~h with 2~wt\% (Fig.~\ref{fig:cumn}). The factors, \NCuMnFacHalf{} and \NCuMnFacTwo, equal $1 + \beta_{\mathrm{Mn}} c_0$, because an immobile, uniform solute scales $\kappa$ uniformly. For the same reason the immortal fractions with reservoir ends, \NCuMnImmWide{} (wide), \NCuMnImmNarrow{} (narrow) and \NCuMnImmNB{} (near-bamboo) at 3~\MAcm, do not depend on the Mn content. This case shows that the solver runs a different host and solute with the same code.

%% file: sections/comparison.tex
\section{Experimental results and comparison with the model}
\label{sec:comparison}

\begin{figure}[t]
\includegraphics[width=\columnwidth]{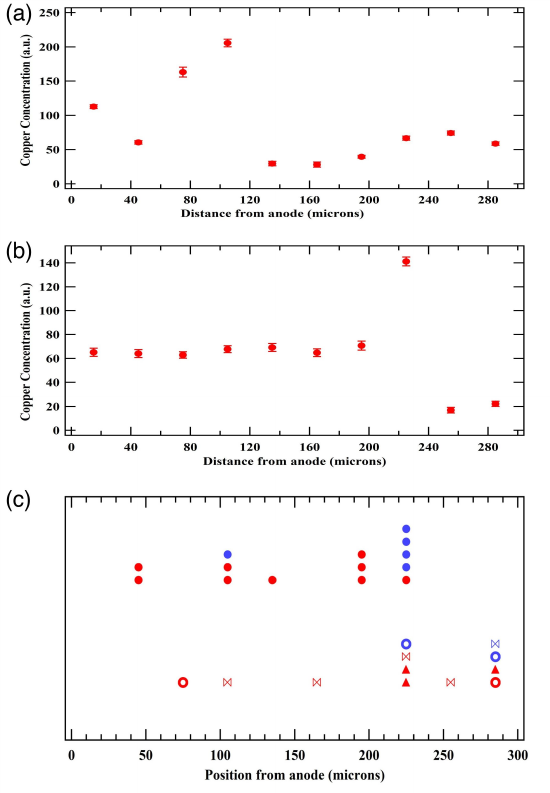}
\caption{Measured Cu profiles and damage sites after stressing at 200~$^{\circ}$C.\cite{dingari2007} (a) Cu concentration (Cu K$\alpha$ EDS intensity, arbitrary units) vs. distance from the anode for wide line w1 (10~\um{} $\times$ 280~\um, Al-2~wt\%~Cu, 2~\MAcm, failed after about 6~days). (b) The same for narrow line n2 (3~\um{} $\times$ 280~\um, Al-0.5~wt\%~Cu, 5~\MAcm, failed after about 11~days). (c) Positions of the Cu maxima (filled circles, upper rows) and of voids (open circles), bright spots (triangles) and extrusions (bow ties) (lower rows) vs. distance from the anode, for the wide lines w1 to w5 (red) and the narrow lines n1 to n6 (blue). The cathode is at 280~\um. Each point in (a) is the mean over an area of about 10~\um{} $\times$ 25~\um; in (b) the areas were centered about 30~\um{} apart and did not reach the neighboring line (Sec.~\ref{sec:experiment}). Original plots, reproduced without change.}
\label{fig:experiment}
\end{figure}

Figure~\ref{fig:experiment} shows a Cu profile of a wide and of a narrow line and summarizes the Cu maxima and damage sites of all eleven lines. Damage, in the form of voids, extrusions and bright spots, was seen in all five wide lines but in only two of the six narrow lines, and most of it lies toward the cathode. In wide lines some damage appeared from the middle of the line toward the anode, but none at the anode end; in narrow lines it was found only toward the cathode. In wide lines the Cu concentration is higher toward the anode and lower toward the cathode, with peaks in three groups at about 40 to 50, 100 to 150 and 190 to 230~\um{} from the anode, most of them in the middle of the line [Figs.~\ref{fig:experiment}(a) and \ref{fig:experiment}(c)]. In four of the six narrow lines the largest Cu concentration lies about 225~\um{} from the anode, $x/L \approx 0.20$, toward the cathode [Figs.~\ref{fig:experiment}(b) and \ref{fig:experiment}(c)].

We compare these observations with simulations at the measured width, length, temperature, current density and Cu content: wide lines at 2~wt\% Cu and 2~\MAcm, and narrow lines at 0.5~wt\% Cu and 5~\MAcm. The Cu profiles are compared at the duration of the tests (E12): about 6~days for the wide and 11~days for the narrow lines, at their own current densities. The Cu equation of the model has no stress term, so the Cu field can be continued past nucleation without any assumption about the stress after it; \NTwelveMaps{} maps per case, and the end condition does not enter because the Cu flux is zero at both ends. The comparison is qualitative and retrospective. The model gives the time and site of the first void nucleation, whereas the experiment imaged damage and measured Cu after failure, on a few lines of each width. Table~\ref{tab:compare} summarizes five checks.

\emph{(1) Damage mostly at the cathode, but not only there.} Narrow lines with blocking ends match this pattern: \NCathNarrowBlk\% of the voids form at the cathode, \NLtHalfNarrowBlk\% in the cathode half and only \NGtNineNarrowBlk\% within $x/L > 0.9$ (Fig.~\ref{fig:location}). Near-bamboo lines show the same tendency, more weakly. Wide lines with blocking ends fail only at the cathode, so the model does not give the interior damage seen in the wide lines. With reservoir ends the sites have no cathode preference.

\emph{(2) Wide lines: Cu build-up toward the anode and Cu peaks in the middle.} At the test duration the model reproduces the build-up, its size and the mid-line maxima, but not their height. After 6~days at 2~\MAcm{} the median ratio of the mean Cu fraction in the last tenth of the line to $c_0$ is \NTwelveWideAnode{} (10th to 90th percentile \NTwelveWideAnodeLo{} to \NTwelveWideAnodeHi), and the largest Cu is at the anode end in \NTwelveWideAnodeMaxPct\% of the maps. Averaged over ten 28~\um{} windows, about the size of the EDS areas, the Cu varies along the line by a median factor of \NTwelveWideContrast{} (\NTwelveWideContrastLo{} to \NTwelveWideContrastHi), as the measured profile of Fig.~\ref{fig:experiment}(a) varies by a factor of several. The largest interior maximum, excluding the first and last tenths, lies in the middle half of the line in \NTwelveWideMidPct\% of the maps, spread much as the three measured groups are. In line w1 a mid-line maximum is higher than the Cu at the anode end [Fig.~\ref{fig:experiment}(a)], whereas in the model the anode end holds the most Cu in almost every map. The build-up takes time: at the nucleation time of the model the anode ratio is still only \NAnodeWideBlkCuTwo{} (E2), and after six hours \NTwelveWideAnodeSixH, because only the dissolved Cu drifts.

\emph{(3) Narrow lines: Cu peaks toward the cathode.} The model reproduces neither these peaks nor the mild depletion measured near the cathode. After 11~days at 5~\MAcm{} the model empties the cathode end of Cu over a median length of \NTwelveNarrowDepUm~\um{} (10th to 90th percentile \NTwelveNarrowDepUmLo{} to \NTwelveNarrowDepUmHi~\um), whereas in line n2 the lowest Cu, about 0.6 times the line mean, is a single area about 25~\um{} from the cathode, and the area at the cathode end is back near the mean [Fig.~\ref{fig:experiment}(b)]. The model therefore moves more Cu out of the cathode end of the narrow lines than was measured. The largest Cu lies at $x/L \approx 0.20$ in four of six measured lines; in the model the largest 28~\um{} window lies at $x/L = 0.1$ to 0.3 in only \NTwelveNarrowPeakCathPct\% of the maps, and in the cathode half in \NTwelveNarrowPeakHalfPct\%. In the model the Cu peaks form where the dissolved Cu flux converges, at boundaries where the Cu diffusivity falls in the direction of the drift, and most lie in the anode half. A line temperature of 225 or 250~$^{\circ}$C, for Joule heating, does not change this (\NTwelveNarrowPeakCathPctTTF\% and \NTwelveNarrowPeakCathPctTFiftyC\%). A Cu pile-up at the depletion front, as in the edge-drift picture of Kim \emph{et al.},\cite{kim1997} would need a mechanism the model lacks, such as finite-rate precipitation at the front.

\emph{(4) Lifetimes.} The narrow lines lasted longer than the wide lines at a current density 2.5 times higher. This is not a like-for-like test: width, Cu content and current density all differ between the two groups, and the measured times are resistance failures, not void nucleation. In the model the wide lines last longer: the medians are \NTWideBlkCuTwoJTwo~h (wide) and \NTNarrowBlkCuHalfJFive~h (narrow) with blocking ends, a factor of \NThesisRatioBlk, and \NThesisRatioRes{} with reservoir ends. The measured failure times are \NGapWideBlk{} (wide) to \NGapNarrowBlk{} (narrow) times the simulated nucleation times with blocking ends (Sec.~\ref{sec:discussion}).

\emph{(5) Which end condition matches (1)?} Blocking ends, through the narrow class and less clearly the near-bamboo class: the cathode-end stress is the largest, and interior divergences next to bamboo grains compete with it. With reservoir ends the end stress is pinned and failures are spread almost evenly along the line. The measured lines ended in wider terminals on the same metal level, not in vias (Sec.~\ref{sec:experiment}), which is closer to a reservoir than to a blocking end. A narrow line joining a wide terminal is not a perfect reservoir, however, because the change in grain structure at the junction is itself a flux divergence. The real ends therefore lie between the two limits, and the better match of blocking ends is consistent with junctions that supplied little Al during the test.

\begin{table}
\caption{Qualitative checks of the model against the measurements.\cite{dingari2007} Wide lines: Al-2~wt\%~Cu, 2~\MAcm; narrow lines: Al-0.5~wt\%~Cu, 5~\MAcm; 200~$^{\circ}$C. Cu profiles at the test durations (E12). B, blocking ends; R, reservoir ends.}
\label{tab:compare}
\begin{ruledtabular}
\begin{tabular}{p{2.6cm}p{3.5cm}p{1.5cm}}
Observation & Model & Verdict \\
\hline
(1) Damage mostly at the cathode, not only there & Narrow (B): \NCathNarrowBlk\% cathode, rest interior; wide (B): cathode only; R: no cathode bias & partly agrees \\
(2) Wide: anode Cu build-up, mid-line peaks & At the test duration: anode build-up, contrast of several, mid-line maxima; anode end always highest & mostly agrees \\
(3) Narrow: Cu peaks toward the cathode, mild cathode depletion & Peaks toward the anode; Cu-free cathode zone of about \NTwelveNarrowDepUm~\um & disagrees \\
(4) Narrow (5~\MAcm) outlives wide (2~\MAcm) & Wide outlives narrow; nucleation times \NGapWideBlk{} to \NGapNarrowBlk{} times shorter than failure times & disagrees (not like-for-like) \\
(5) End condition for (1) & Blocking & B partly, R no \\
\end{tabular}
\end{ruledtabular}
\end{table}

%% file: sections/discussion.tex
\section{Discussion}
\label{sec:discussion}

\subsection{When Cu redistribution matters}\label{sec:disc:which}

With blocking ends, nucleation at the cathode needs the stress to rise by $\sigma_{\mathrm{crit}} - \sigma_{\mathrm{res}}$ over a layer of order $\sigma_{\mathrm{crit}}/G$, which at 2 to 5~\MAcm{} is one or two grains long. This takes $t_{\mathrm{nuc}} \approx \pi\sigma_{\mathrm{crit}}^{2}/(4G^{2}\kappa_{\mathrm{eff}})$ (Sec.~\ref{sec:res:e1}), and the Cu field enters through $\kappa_{\mathrm{eff}}$ in that layer. How much it matters depends on how much of the Cu can move. Three treatments, run on the same maps, span this (Sec.~\ref{sec:res:cutransport}). With the Cu immobile, the median nucleation time is the static value, $1 + \beta c_0$ times the Cu-free one. With only the dissolved Cu mobile, the main model, it is \NSevenRatioBlkWNCuTwoMin{} to \NSevenRatioBlkWNCuHalfMax{} times shorter, at both Cu contents and every current density. With all Cu mobile, the assumption of earlier solute-coupled models, it is shorter again, by \NTenRatioTotBlkWNCuHalfMin{} to \NTenRatioTotBlkWNCuHalfMax{} at 0.5~wt\% and \NTenRatioTotBlkWNCuTwoMin{} to \NTenRatioTotBlkWNCuTwoMax{} at 2~wt\%. The last factor grows with the Cu content because the excess above the solubility, which the all-mobile model lets drift, grows with it. None of the factors depends on $j$. A scaling argument is consistent with this: the layer length scales as $1/j$, and the times for Cu to drift ($v \propto j$) or diffuse across it scale as $j^{-2}$, like $t_{\mathrm{nuc}}$, so the fraction of Cu lost from the layer at nucleation does not depend on $j$. The exponent $n$ therefore stays at 2.

Claim 1 of Sec.~\ref{sec:intro} is thus a tested, conditional result. In cathode-dominated populations with blocking ends, the evolving Cu field changes the median nucleation time by a modest factor and the share of cathode sites in narrow lines, but not $n$. It changes the statistics most where the microstructure alone gives little stress, in wide lines with reservoir ends. There, the dissolved Cu moves while the precipitate stays, so the total Cu develops sharp steps that diffusion does not smooth, and these steps become the flux divergences at which the lines fail. Copper then removes the immortality of these lines instead of adding to it, and brings their exponent from about 3.6 down to about 2.2. Neither immobile Cu nor fully mobile Cu gives this result, so it depends on the solubility limit. It also needs the precipitate to block Al diffusion, but only partly: a quarter of the precipitate blocking is enough (E11). Its size depends on how fast precipitates dissolve and grow, which the model takes as instantaneous; finite rates would smooth the steps. The incubation-then-drift sequence seen in experiments\cite{hu1993,proost2000,kao2001} would matter more for the slower stages that follow nucleation, which the model does not include.

\emph{Which Cu blocks?} The Cu-step failures need some of the precipitate to block Al diffusion, and the literature does not settle whether it does. Frear \emph{et al.} list five proposed mechanisms for the Cu effect and find no agreement among them.\cite{frear1991} Rosenberg attributed it to binding between Cu solute and grain-boundary defects.\cite{rosenberg1972,wang2024} In the line of Wang \emph{et al.}, where all the Cu was dissolved, Al grain-boundary diffusion was effectively stopped above about 0.16~at.\% Cu.\cite{wang2024} Hu \emph{et al.}\cite{hu1993} infer that Cu electromigration along grain boundaries, not the dissolution of Al$_{2}$Cu precipitates, controls the kinetics of damage formation, and in submicron bamboo lines precipitates formed before stressing lowered the lifetime.\cite{theiss1997} Frear \emph{et al.} propose that a thin, coherent film of $\theta$ phase or a precursor on the grain boundaries retards diffusion, and note that the incoherent interfaces of discrete precipitates could still be fast paths.\cite{frear1991} Kim, Morris and Lee\cite{kim1997} model the growth of the precipitate-free zone at the cathode end; their analysis of earlier tests suggests that drift begins almost as soon as this zone exceeds the Blech length, and so that grain-boundary Al$_{2}$Cu precipitates are essential to slowing Al electromigration.

Three observations favor some blocking by precipitate ($f > 0$) in polycrystalline lines at 200~$^{\circ}$C. First, the solubility at 200~$^{\circ}$C, 0.10~wt\% (0.04~at.\%), is a quarter of the content at which Wang \emph{et al.} find Al grain-boundary diffusion blocked, yet Knowlton's Al-1~wt\% Cu diffusivity at this temperature is 50 times below that of pure Al.\cite{knowltonthesis} Second, with $f = 0$ the model gives almost the same median nucleation time at 2 and 0.5~wt\% Cu, a ratio of \NElevenCuRatioZeroMin{} to \NElevenCuRatioZeroMax{} over the line classes and end conditions (geometric mean over $j$), because both contents exceed the solubility; every $f \ge 0.25$ gives \NElevenCuRatioPartMin{} to \NElevenCuRatioPartMax. The incubation time of Kim \emph{et al.} also grows with the initial Cu content, and Theiss \emph{et al.} note that, unlike in their bamboo lines, raising the Cu content from 0.5 to 0.9~wt\% does change the lifetime of polygranular lines.\cite{theiss1997} Third, in the wide Al-2~wt\% Cu lines of Sec.~\ref{sec:comparison} damage lay close to, and on the cathode side of, several Cu maxima.\cite{dingari2007} There the Cu rises in the direction of Al drift, so if Cu-rich material blocks, the Al flux converges and builds compression, consistent with the extrusions and hillocks found in these lines; voids, which in the model form where the Cu falls in the drift direction (Sec.~\ref{sec:res:e3}), would not be explained this way. These lines hold on average 20 times the soluble Cu, so with $f = 0$ the blocking would be the same on both sides of most maxima and would not localize damage at them. Each of the three is indirect, and none fixes $f$. We therefore state the result conditionally: wide reservoir-end lines fail at Cu steps for every $f \ge 0.25$ (E11) and stay largely immortal only if the precipitate does not block at all. The main model takes $f = 1$ because $\beta$ was calibrated on the total Cu of Knowlton's Al-1~wt\% Cu; partial blocking lowers the median nucleation time roughly in proportion to the unblocked fraction, so the lifetimes of Cu-rich lines are uncertain by up to a factor of about three from $f$ alone if $f \ge 0.25$, and by much more if the precipitate does not block at all: with $f = 0$ the median nucleation time of the wide blocking-end lines at 2~wt\% Cu and 3~\MAcm{} falls from \NTenTTotWideBlkCuTwoJThree~h to \NTenTDisWideBlkCuTwoJThree~h.

\subsection{The lifetime gap: nucleation against resistance failure}

The simulated nucleation times are \NGapWideBlk{} to \NGapNarrowBlk{} times shorter than the measured failure times (Sec.~\ref{sec:comparison}). The two quantities are different: the model stops when a void nucleates, whereas the experiment recorded resistance failure, which follows void growth, and nucleation-limited and growth-limited failure scale differently ($n = 2$ against $n = 1$).\cite{park1999,lloyd2007} The inputs are also uncertain: the cluster diffusivity of Al-Cu at 200~$^{\circ}$C is quoted over more than an order of magnitude,\cite{knowltonthesis} the Cu diffusivity prefactor is uncertain by a factor of about 40,\cite{kao2001} and $\beta$ inherits Knowlton's fit of the Al-Cu diffusivity to failure times. E8 changes one input at a time for the uniform line with 0.5~wt\% Cu and blocking ends (Table~\ref{tab:sens}). Within the base ranges each change alters $t_{\mathrm{nuc}}$ by a factor between \NEightSingleMin{} and \NEightSingleMax, the same at 2 and 5~\MAcm{} within \NEightJDiffMax\%. Raising $\sigma_{\mathrm{crit}}$ to 1~GPa, the upper end of its quoted range,\cite{knowltonthesis} lengthens it by \NEightSigCritGPa, much less than $t_{\mathrm{nuc}} \propto \sigma_{\mathrm{crit}}^{2}$ would give, because Cu leaves the cathode layer during the longer build-up. Only with every input at its long-life end (\NEightCombined, or \NEightCombinedGPa{} with $\sigma_{\mathrm{crit}}$ at 1~GPa) does $t_{\mathrm{nuc}}$ approach the gap measured for the wide lines. The most likely explanation of the gap is therefore the difference between nucleation and resistance failure.

\begin{table}
\caption{One-at-a-time sensitivity of the nucleation time (E8): uniform line, 0.5~wt\% Cu, blocking ends, 200~$^{\circ}$C. Entries are $t_{\mathrm{nuc}}$ relative to the baseline (\NEightBaseJTwo~h at 2~\MAcm{} and \NEightBaseJFive~h at 5~\MAcm); the values at 2~\MAcm{} are shown, and those at 5~\MAcm{} agree within \NEightJDiffMax\%. The last two rows set every input at the end of its range that lengthens $t_{\mathrm{nuc}}$, with $\sigma_{\mathrm{crit}}$ doubled or at 1~GPa.}
\label{tab:sens}
\begin{ruledtabular}
\begin{tabular}{lcc}
Input & Lower setting & Upper setting \\
\hline
$\sigma_{\mathrm{crit}}$ ($\times0.5$, $\times2$) & \NEightSigCritHalf & \NEightSigCritTwo \\
$\sigma_{\mathrm{crit}}$ (600~MPa, 1~GPa) & \NEightSigCritFour & \NEightSigCritGPa \\
$D_{0,\mathrm{gb}}$ ($\times0.1$, $\times10$) & \NEightDzeroTenth & \NEightDzeroTen \\
$E_{a,\mathrm{gb}}$ ($-0.05$, $+0.05$~eV) & \NEightEaMinus & \NEightEaPlus \\
$\beta$ ($\times0.5$, $\times2$) & \NEightBetaHalf & \NEightBetaTwo \\
$\Zs_{\mathrm{Al}}$ ($\times0.75$, $\times1.25$) & \NEightZLow & \NEightZHigh \\
All long-life, $\sigma_{\mathrm{crit}}\times2$ & \multicolumn{2}{c}{\NEightCombined} \\
All long-life, $\sigma_{\mathrm{crit}}$ 1~GPa & \multicolumn{2}{c}{\NEightCombinedGPa} \\
\end{tabular}
\end{ruledtabular}
\end{table}

\subsection{Extrapolation from accelerated tests}

Rule-based and physics-based electromigration checks share one step: lifetimes measured at high $T$ and $j$ are carried to use conditions with the $n$ and $E_a$ fitted at the test conditions.\cite{sapatnekar2026} E9 tests that step on the simulated populations (Sec.~\ref{sec:res:e9}). Two kinds of error appear. The first is a change of failure mode between test and use conditions. At 1~\MAcm{} part of the near-bamboo population outlives the prediction by orders of magnitude, and Cu-free wide lines with reservoir ends become largely immortal, which no finite Black's-law prediction represents. These errors are conservative, because the mechanisms that appear at low $j$, blocking by bamboo grains and steady-state immortality, lengthen the life. An exponent well above 2 at the test conditions is their warning sign: in our populations it marks interior failure or a bimodal population, not the mode that governs at low $j$. Al electrodes in SiC contact holes give $n = 3.0$ to 3.4 at 300 to 450~$^{\circ}$C and are extrapolated over two decades of $j$ with that exponent;\cite{tanimoto2010} the authors attribute the high exponent to the three-dimensional via geometry, which our one-dimensional model cannot test, but our results show why such an extrapolation needs a check that the failure mode does not change.

The second error has no such warning sign. With 2~wt\% Cu the failure mode is the same at test and use conditions, the fitted exponent is close to 2, and the extrapolation still overestimates the median nucleation time, by a factor of up to \NNineRatioMax{} at 150~$^{\circ}$C. The cause is the solvus. The fraction of the Cu that is dissolved, and so able to leave the cathode layer, rises with temperature, so the apparent activation energy is higher between 250 and 300~$^{\circ}$C (\NNineEaLocCuTwoHi~eV) than between 150 and 200~$^{\circ}$C (\NNineEaLocCuTwoLo~eV), and an activation energy fitted at the test temperatures predicts too long a life at the use temperature. The error grows with the distance in temperature and with the Cu content above the solvus; at 0 and 0.5~wt\% Cu, which is almost fully dissolved at 300~$^{\circ}$C, it is within 10\% in the cathode-limited lines. Accelerated tests of alloys with Cu above the solvus therefore need either test temperatures on the same side of the solvus as the use temperature, or an activation energy that accounts for the solvus. These statements rest on nucleation times. Void growth, which the model omits, has $n$ near 1 (Ref.~\onlinecite{park1999}) and would change the numbers, but not the argument.

\subsection{Limits of the model}

The model is one-dimensional. Grain-boundary networks, triple junctions and the line width enter only through the segment classification and the diffusivity multipliers, and the transverse stress state is folded into the effective modulus. Void growth, the resistance change, and the stress dependence of the diffusivity are not included, and neither are Joule heating or thermal stress beyond the uniform $\sigma_{\mathrm{res}}$. The random-map runs set $\sigma_{\mathrm{res}} = 0$. The measured lines were passivated, so they carried a tensile thermal stress at the test temperature; a positive $\sigma_{\mathrm{res}}$ lowers the stress rise $\sigma_{\mathrm{crit}} - \sigma_{\mathrm{res}}$ needed for nucleation and shortens the simulated times, which would widen the lifetime gap discussed above. The Cu blocking law is linear in $c$ and calibrated at one Cu content; bamboo Cu diffusion is scaled from the grain-boundary value without a direct measurement. Precipitation is treated as instantaneous local equilibrium: only Cu below the solubility moves, and the excess is an immobile precipitate that dissolves or grows at once as the dissolved flux requires. Real $\theta$-Al$_{2}$Cu precipitates dissolve and coarsen at finite rates and can themselves migrate under current,\cite{witt2003} so the cap is a limiting case of this step, not a description of it. The cap uses the bulk solubility, while Cu moves along grain boundaries, where segregation raises the local Cu content above the lattice value. The mobile grain-boundary inventory in equilibrium with $\theta$ may therefore exceed $c_{\mathrm{sol}}$; the immobile and all-mobile limits bracket this case, and the capped results should be read between them. E7 (immobile Cu) and E3\textsubscript{mobile} (all Cu mobile) are the opposite limits of Cu transport (Sec.~\ref{sec:res:cutransport}); they are limiting scenarios, not proven bounds on every observable, and finite-rate precipitation would smooth the Cu steps that the cap produces. The local Cu content of lines that nucleate a void stays below \NCuMaxNucPct~at.\%, under the composition of $\theta$. In short blocking-end lines held at steady state, however, the precipitate piles up in a few anode cells, which is why the Blech-length study E4 uses the all-mobile model (Sec.~\ref{sec:methods}). Which Cu blocks Al diffusion is the least certain input of the model; its consequences are discussed in Sec.~\ref{sec:disc:which}. The Cu flux is zero at both ends even with reservoir ends; a pad that exchanged Cu with the line could weaken the Cu steps next to it. The Cu transport equation is a reduced drift-diffusion model with a constant effective charge, without stress-driven or nonideal chemical-potential terms. The Cu inputs were varied on populations in E13 (\NThirteenMaps{} maps, 3~\MAcm). The shortening of the median nucleation time by depletion, \NThirteenBaseMin{} to \NThirteenBaseMax{} in cathode-limited lines with the cited inputs, is \NThirteenZMin{} to \NThirteenZHiMax{} over the measured range of $Z^{*}_{\mathrm{Cu}}$ and \NThirteenSolTwoMin{} to \NThirteenSolTwoMax{} with twice the solubility, but it depends strongly on $D_{0,\mathrm{Cu}}$: about 1.0 if Cu diffuses 40 times slower and \NThirteenDFastMin{} to \NThirteenDFastMax{} if it diffuses 40 times faster. The depletion factors of this paper are therefore set mainly by the Cu diffusivity, which is uncertain by that factor of 40 either way. The earlier failure of wide reservoir-end lines at Cu steps persists over these inputs, though its size does not: at 3~\MAcm{} the immobile-Cu median exceeds the capped one by a factor of \NThirteenResBaseMin{} to \NThirteenResBaseMax{} with the cited inputs and by only \NThirteenResDSlowMax{} with the slowest Cu. E13 did not test the low current densities at which the Cu-free lines become immortal. The extrapolation error of Sec.~\ref{sec:res:e9} was not recomputed for these inputs; it depends on them through the dissolved fraction. The compressive relief is a numerical device; with the solubility cap no result depends on it (Sec.~\ref{sec:res:e3}). The two end conditions are idealizations of vias and pads, and the comparison favors neither cleanly. The solver is verified against analytic solutions and conservation checks. Within these limits the results are internally consistent, and every physical input apart from the stated assumptions below is cited. The relief parameters $k_r$ and $w$ and the multiplier spread $\sigma_{\ln}$ are stated assumptions, not cited inputs (Table~\ref{tab:params}). The immortality of the Cu-free wide lines with reservoir ends, which Cu removes, rests on the $\kappa$ contrast set by $\sigma_{\ln}$ without misorientation spread or triple junctions; a wider contrast could make Cu-free lines fail as well. The model also omits Joule heating, which was larger in the narrow lines at 5~\MAcm, and uses a grain size measured for Al-Cu on TiN because none was found for TiW;\cite{atakov1993} either could move the Cu peaks of the narrow lines. The uniform initial Cu field of the model matches the Cu profile measured after the 24~h anneal at 200~$^{\circ}$C (Sec.~\ref{sec:experiment}). No input was tuned to the measurements. The model predicts void nucleation, not resistance failure, so in its present form it should not be used as a quantitative predictor of measured failure times. Two results have no isolated mechanism and are model predictions still to be tested: the exponent slightly above 2 of Cu-free near-bamboo lines with blocking ends (\NBnNBBlkCuZeroJFourSix{} to \NBnNBBlkCuZeroJTwoFive) and the exponent of Cu-free wide lines with reservoir ends at 4 to 6~\MAcm{} (\NBnWideResCuZeroJFourSix).

%% file: sections/conclusion.tex
\section{Conclusion}
\label{sec:conclusion}

We coupled an evolving Cu field to the Korhonen stress equation on populations of random grain-segment maps, solved it for wide, narrow and near-bamboo Al-Cu lines with an open, verified solver, let only the Cu below its solubility move, and compared this with the limiting cases of immobile and fully mobile Cu. Five results follow.

First, with blocking ends and the cited parameters, nucleation happens in a cathode layer about one grain long. Depletion of that layer by the dissolved Cu shortens the median nucleation time by a factor of \NSevenRatioBlkWNCuTwoMin{} to \NSevenRatioBlkWNCuHalfMax{} compared with immobile Cu, at every current density and both Cu contents. This factor is set mainly by the Cu diffusivity: over its factor-of-40 uncertainty it ranges from about 1.0 to \NThirteenDFastMax{} (E13). Treating all Cu as mobile overstates this by a further \NTenRatioTotBlkWNCuHalfMin{} to \NTenRatioTotBlkWNCuTwoMax-fold, growing with the Cu content. Net of depletion, 0.5 and 2~wt\% Cu raise the median nucleation time about \NCuFacWideBlkHalf- and \NCuFacWideBlkTwo-fold.

Second, with \NEthreeMaps{} maps per condition the nucleation-time distributions are narrow ($\sigma_{\ln} \le \NSigBlkWNMax$) for wide and narrow lines with blocking ends and bimodal for near-bamboo lines at 1~\MAcm. Wide lines with reservoir ends are largely immortal at low current density without Cu, but with Cu every such line fails: the precipitate does not move, so the total Cu develops sharp steps that act as flux divergences. This holds whenever at least a quarter of the precipitate blocks Al diffusion; partial blocking lowers the Cu benefit in proportion but keeps these failures. The rise of lifetime with Cu content and the damage found next to measured Cu maxima favor such blocking, though neither fixes its size. With blocking ends the Blech threshold does not depend on the microstructure or the Cu content.

Third, the apparent current-density exponent stays near 2 (\NBnWideBlkMin{} to \NBnNarrowBlkMax) in wide and narrow lines with blocking ends, with no, immobile, capped or fully mobile Cu, in agreement with earlier nucleation models of Al-Cu. Near-bamboo lines give \NBnNBJgeMin{} to \NBnNBJgeMax{} above 1~\MAcm, independent of the Cu content; the rise with Cu content found with fully mobile Cu does not survive the solubility limit. Larger values arise from bimodality and from immortality or its approach, as Knowlton found without solute. We do not report $n$ where the immortal fraction reaches \NImmGate.

Fourth, over 150 to 300~$^{\circ}$C the populations with blocking ends have an apparent activation energy of \NNineEaBlkMin{} to \NNineEaBlkMax~eV, but with Cu above its solubility the lifetime is not Arrhenius, because the dissolved fraction rises along the solvus. A Black's-law fit at \NNineAccTlo{} to \NNineAccThi~$^{\circ}$C and 4 to 6~\MAcm{} predicts the median at 150 to 200~$^{\circ}$C and 1 to 2~\MAcm{} within about 10\% for cathode-limited lines with 0 or 0.5~wt\% Cu, but overestimates it up to \NNineRatioMax-fold with 2~wt\% Cu, with no change of failure mode and an exponent near 2. Where the failure mode changes, at low $j$ in near-bamboo lines and Cu-free wide lines with reservoir ends, the extrapolation underestimates the lifetime. Within this nucleation model, a fixed exponent and activation energy carried from test to use conditions are therefore reliable only where neither the failure mode nor the dissolved Cu fraction changes between them; the size of the error depends on the Cu inputs and awaits an experimental test.

Fifth, the model is qualitatively consistent with the measured cathode-side damage with some interior sites in narrow lines and, when the Cu field is followed to the test duration, with the anode Cu build-up and mid-line Cu peaks in wide lines. It does not reproduce the Cu peaks toward the cathode in narrow lines, where it also removes more Cu from the cathode end than was measured, the longer lifetimes of the narrow lines, or the magnitude of the lifetimes. Void growth, finite-rate precipitation and a measured grain structure for the tested lines are the next steps that these disagreements point to.

%% file: sections/availability.tex
\section*{Author Declarations}

\subsection*{Conflict of Interest}
The authors have no conflicts to disclose.

\subsection*{Author Contributions}
\textbf{Narahara Chari Dingari:} Conceptualization, Methodology, Software, Formal analysis, Investigation, Data curation, Visualization, Writing -- original draft.\\
\textbf{David R. Heskett:} Validation, Writing -- review \& editing.

\subsection*{AI-use disclosure}
The use of AI tools in this work, including in the design of the experiments, the choice of statistical methods and the running of the simulations, is described in Sec.~\ref{sec:ai}.

\section*{Data Availability}
The simulation code, the simulation results and the data behind every figure are kept in a GitHub repository (\url{https://github.com/ncdingari/solute-aware-korhonen-em}), which will be made public and archived in Zenodo, with a DOI, at publication. Until then they are available from the corresponding author upon reasonable request, including to the referees.